\documentclass[12pt,reqno]{amsart}
\usepackage{amsthm}
\usepackage{amsmath}
\usepackage[foot]{amsaddr}
\usepackage{latexsym}
\usepackage{amsfonts}
\usepackage{amssymb}
\usepackage{color}
\usepackage{bbm,dsfont}
\usepackage{graphicx}
\usepackage{subfigure}
\usepackage{mathrsfs}
\usepackage{mathbbol}
\usepackage{enumerate}
\usepackage{MnSymbol}
\usepackage{bbding}
\usepackage{multirow}
\usepackage{array}
\usepackage{makecell}
\usepackage[table]{xcolor}
\usepackage{lscape}

\newtheorem{proposition}{Proposition}
\newtheorem{proposition?}{Proposition?}
\newtheorem{theorem}{Theorem}

\theoremstyle{definition}

\newtheorem{example}{Example}
\newtheorem{definition}{Definition}

\definecolor{cviolet}{RGB}{148,0,211} 
\definecolor{cblue}{RGB}{0,127,255} 
\definecolor{ccyan}{RGB}{0,255,255} 
\definecolor{cgreen}{RGB}{0,255,0}
\definecolor{cyellow}{RGB}{253,218,13} 
\definecolor{corange}{RGB}{255,127,0}
\definecolor{cred}{RGB}{255,0,0}

\newcommand{\cplus}{\cellcolor{gray!10}}
\newcommand{\csum}{\cellcolor{teal!10}}

\newcommand{\boldv}{{\vrule width 1.2pt}} 
\newcommand{\boldh}{\Xhline{1.2pt}} 

\newcommand{\ce}{\cellcolor{cviolet!70}e} 

\newcommand{\cg}{\cellcolor{cgreen!60}g}
\newcommand{\ch}{\cellcolor{cyellow!70}h}
\newcommand{\ci}{\cellcolor{corange!70}i}
\newcommand{\cj}{\cellcolor{cred!70}j}

\newcommand{\ccv}[1]{\cellcolor{cviolet!70}#1}
\newcommand{\ccb}[1]{\cellcolor{cblue!70}#1}
\newcommand{\ccc}[1]{\cellcolor{ccyan!70}#1}
\newcommand{\ccg}[1]{\cellcolor{cgreen!70}#1}
\newcommand{\ccy}[1]{\cellcolor{cyellow!70}#1}
\newcommand{\cco}[1]{\cellcolor{corange!70}#1}
\newcommand{\ccr}[1]{\cellcolor{cred!60}#1}

\newcommand{\ccone}{\cellcolor{black!20}$\eone$}
\newcommand{\ccnot}{\cellcolor{white!10}$\diamond$}

\newcommand{\real}{\mathbb R} 
\newcommand{\complex}{\mathbb C} 
\newcommand{\half}{\tfrac{1}{2}} 

\newcommand{\hi}{\mathcal{H}} 
\newcommand{\eh}{\mathcal{E(H)}} 
\newcommand{\ip}[2]{\left\langle\,#1\,|\,#2\,\right\rangle} 
\newcommand{\id}{\mathbbm{1}} 

\newcommand{\eone}{\mathfrak{1}}
\newcommand{\enul}{\mathfrak{0}}
\newcommand{\ea}{\mathcal{E}}

\newcommand{\sea}[1]{\mathcal{S}_{#1}} 
\newcommand{\dea}[1]{\mathcal{D}_{#1}} 
\newcommand{\pea}[1]{\mathcal{P}_{#1}}
\newcommand{\emix}[1]{\mathcal{M}_{#1}} 
\newcommand{\scale}[1]{\mathcal{S}_{#1}}

\newcommand{\vx}{\vec{x}} 
\newcommand{\vy}{\vec{y}} 

\begin{document}

\title{Finite (quantum) effect algebras}
\author[]{Stan Gudder}
\address[S.G.]{Department of Mathematics, University of Denver, USA}
\author[]{Teiko Heinosaari}
\address[T.H.]{Faculty of Information Technology, University of Jyväskylä, Finland.}

\begin{abstract}
We investigate finite effect algebras and their classification.
We show that an effect algebra with $n$ elements has at least $n-2$ and at most $(n-1)(n-2)/2$ nontrivial defined sums. 
We characterize finite effect algebras with these minimal and maximal number of defined sums.
The latter effect algebras are scale effect algebras (i.e., subalgebras of [0,1]), and only those. We prove that there is exactly one scale effect algebra with $n$ elements for every integer $n \geq 2$.
We show that a finite effect algebra is quantum effect algebra (i.e. a subeffect algebra of the standard quantum effect algebra) if and only if it has a finite set of order-determining states. Among effect algebras with 2-6 elements, we identify all quantum effect algebras.
\end{abstract}

\maketitle

.

\section{Introduction}

The basic setting of a physical experiment consists of states and measurements, and their operational analysis gives a solid starting point for studies on the foundations \cite{SEO83}.
Hence, there are two main branches of axiomatic approaches to quantum theory; either one starts from states, or from measurements.
An algebraic structure called an effect algebra \cite{FoBe94} is one possible starting point to investigate the mathematical structure of quantum measurements, or more precisely, possible events that are constituents of measurements. 
Effect algebras, equivalently introduced also as D-posets \cite{kopka1994d},  have been investigated from various viewpoints \cite{gudder1997effect, NTQS00, GuGr02} and are still under active research.

Previous studies have shown that if an effect algebra has additional structure so that it is a convex effect algebra, then it is isomorphic to the effect algebra arising from a state space that is a convex subset of a vector space \cite{GuPu98,Gudder99, GuPuBuBe99}.
In that way, convex effect algebras are closely linked to the state centric approach of general probabilistic theories \cite{plavala2023general}.
In the current work, our focus is on effect algebras that have only a finite number of elements.
In particular, these effect algebras are not convex effect algebras and therefore they do not have a representation theorem that would connect them to a linear structure. 
Some structural results on finite effect algebras have been reported in \cite{FoGr00, rievcanova2001proper, Jenca03,binczak2023matrix}.
Our approach differs from these earlier investigations by taking the number of defined sums as the main characterizing feature of a finite effect algebra. 
Further, we pay special attention to a question of when an effect algebra can be represented as a finite collection of quantum effects, or when that is not possible.
A quantum effect algebra is an effect subalgebra of the standard quantum effect algebra, which consists of all positive linear operators on a Hilbert space bounded by the identity operator.
We are hence investigating the question of what are the finite quantum effect algebras.

Our motivation for the current investigation is manyfold. 
Firstly, by examining finite effect algebras we can learn what kind of structures are possible e.g. when available measurements have some restrictions, or events are limited in some way.
Secondly, by classifying all the effect algebras of small order we can find simple structures that are not quantum but in any case describe some kind of events, in the abstract sense of effect algebras. 
This sheds light on the question of what mathematical structures are logically possible but not realizable in a quantum physical setup.  
Thirdly, effect algebras provide a potential test bed for examining the distinctions between classical and quantum structures.
Consequently, it is interesting to see whether there are differences in finite effect algebra structures or not.

The essence of an effect algebra structure is the fact that effect sum is not a binary operation but a partial binary operation.
We show that an effect algebra with $n$ elements has at least $n-2$ and at most ${\half(n-1)(n-2)}$ nontrivial defined sums (i.e. those where both elements are different than $\enul$ and $\eone$).
We give a full characterization of finite effect algebras with these minimal and maximal number of defined sums.
We show that the first ones, called sparse effect algebras, are all valid effect algebra structures.
One of our main results is that the effect algebras with a maximal number of defined sums are scale effect algebras (i.e., subalgebras of $[0,1]$), and only those.
We prove that there is exactly one scale effect algebra with $n$ elements for every integer $n\geq 2$.

We introduce the illustrative tool of sum table, and demonstrate its use by giving a complete presentation of all effect algebras with 2 -- 6 elements.
There are in total 19 of such effect algebras. We remark that the number of all effect algebras with 2 -- 8 elements have been previously given in \cite{binczak2023matrix} and our list does not provide new information on that, although the representation via sum tables makes the list easily accessible.
The novelty of our approach is the classification of them into those that have a description as quantum effect algebras and those that do not; see Table \ref{table1}.
We also investigate their state spaces and single out all which have separating state spaces and order-determining state spaces.

Using states of an effect algebra as a tool, we are able to prove two important structural results.
Firstly, we show that an effect algebra has a finite set of order-determining states if and only if it is a fuzzy set effect algebra.
Secondly, if an effect algebra is finite, then a finite set of order-determining states is equivalent of being a quantum effect algebra.
In particular, all finite quantum effect algebras are fuzzy set effect algebras.
We complement this result by demonstrating that the dimensions in representations in terms of fuzzy sets and quantum effects may be different.

In the end of the paper, we discuss composing of effect algebras into a larger effect algebras. 
We demonstrate that some of the presented finite effect algebras are compositions, while others cannot be decomposed into components.
We also prove that a composition of two fuzzy set effect algebras is a fuzzy set effect algebra.
In particular, a composition of two finite quantum effect algebras is again a finite quantum effect algebra.

\begin{table}[htbp]
\centering
\caption{Number of effect algebras with 2--6 elements.}
\label{table1}
\begin{tabular}{|c|c|c|c|c|}
\hline
\textbf{\# elements} & \textbf{\# effect algebras} & \textbf{quantum} & \textbf{non-quantum} \\ 
\boldh
2 & 1 & 1 & 0 \\ \hline
3 & 1 & 1  & 0  \\ \hline
4 & 3 & 2  & 1 \\ \hline
5 & 4 & 2  & 2 \\ \hline
6 & 10 & 4 & 6  \\ \hline
\end{tabular}
\end{table}

\section{Effect algebras}

\subsection{Basic definitions and standard examples}

In this subsection we recall the basic definitions and the standard examples relevant to our investigation.
For the reader's convenience, the proofs of some elementary well-known facts that are repeatedly used are given in the Appendix A.

An \emph{effect algebra} is a set $\ea$ with two special elements $\enul,\eone\in\ea$ and a partial binary operation $\oplus$ (i.e. binary operation that need not be defined for all pairs of elements) satisfying the following conditions:
\begin{itemize}
\item[(EA1)] $e \oplus f = f \oplus e$ 
\item[(EA2)] $e \oplus (f \oplus g) = (e \oplus f) \oplus g$ 
\item[(EA3)] for every $e$ there exists a unique element $e^\perp$ such that $e \oplus e^\bot = \eone$,
\item[(EA4)] if $e \oplus \eone$ is defined, then $e=\enul$.
\end{itemize}
An element of an effect algebra is called an \emph{effect} and the operation $\oplus$ is referred to as effect sum, or just sum.
Unless confusion threatens, we say that $\ea$ is an effect algebra when we really mean that $(\ea,\oplus,\enul,\eone)$ is an effect algebra.

The conditions (EA1) and (EA2) declare the commutativity and associativity of $\oplus$, and they are meant to contain the requirement that the left hand side of the equation is defined if and only if the right hand side is defined. 
The same convention is used in all equations; when we e.g. write $e \oplus f =g$ we mean that $e \oplus f$ is defined and that $e \oplus f =g$.

The condition (EA3) declares that every effect $e$ has a unique \emph{complement} effect $e^{\bot}$.
It is possible that $e=e^\bot$, in which case we say that $e$ is \emph{self-complementary}. 
The conditions (EA3) and (EA4) together imply that $\eone$ is the complement of $\enul$.

For later use, we recall some basic properties of effect algebras.
The following properties follow from (EA1)--(EA4) and their proofs are given in the appendix for completeness.
\begin{itemize}
\item[(EA-a)] if $e \oplus f = e \oplus g$, then $f=g$ (\emph{the cancellation law})
\item[(EA-b)] $e \oplus \enul = e$
\item[(EA-c)] if $e \oplus f = e$, then $f=\enul$
\item[(EA-d)] if $e \oplus f = \enul$, then $e=f=\enul$
\end{itemize}

An effect algebra has an intrinsic partial ordering; we denote $e \geq f$ if there exists $g$ such that $e=f \oplus g$.
The effect $g$, if it exists, is uniquely defined by the cancellation law and we denote $g=e \ominus f$. 
The proof of the following basic fact is given in the appendix.
\begin{itemize}
\item[(EA-e)] $e \ominus f = (e^\bot \oplus f)^\bot$
\end{itemize}
We further denote $e>f$ if $e \geq f$ and $e \neq f$.

We recall the usual examples of effect algebras.

\begin{example}(\emph{Standard scale effect algebra})
The interval $[0,1]\subset\real$ is an effect algebra where  $x\oplus y$ is defined whenever $x+y \leq 1$ and in this case $x\oplus y = x+y$. 
This is called the \emph{standard scale effect algebra}.
\end{example}

\begin{example}(\emph{Standard fuzzy set effect algebra})
Let $n$ be a positive integer.
The set $[0,1]^n$ is an effect algebra when $\vx \oplus \vy$ is defined if and only if $x_i + y_i \leq 1$ for all $i=1,\ldots,n$ and then $\vx \oplus \vy$ is the componentwise sum of the vectors.
The elements of $[0,1]^n$ can be interpreted as fuzzy sets on $\{1,\ldots,n\}$. 
We call $[0,1]^n$ the \emph{standard fuzzy set effect algebra} of dimension $n$.
The standard scale effect algebra corresponds to $n=1$.
\end{example}

\begin{example}(\emph{Standard quantum effect algebra})
Let $\hi$ be a $d$-dimensional Hilbert space. We denote by $\eh$ the set of all bounded linear operators $E$ satisfying the operator inequalities $ 0 \leq E \leq \id$. We define $E\oplus F = E+F$ whenever $E+F \leq \id$.
We call $\eh$ the \emph{standard quantum effect algebra} of dimension $d$.
\end{example}

In our later investigation, a special role will be on those effect algebras that are subalgebras of these three standard cases.
However, we are interested solely in the effect algebra structure and other details are not important for us. 
An effect algebra isomorphisms preserve all the effect algebra operations so two isomorphic effect algebras are essentially the same.
We recall that given two effect algebras $\ea_1$ and $\ea_2$, a map $\Theta: \ea_1 \to \ea_2$ is a \emph{morphism} if $\Theta(\eone) = \eone$ and $\Theta(e) \oplus \Theta(f)$ is defined when $e \oplus f$ is defined and then $\Theta( e \oplus f ) = \Theta(e) \oplus \Theta(f)$. 
A morphism is a \emph{monomorphism} if $\Theta(e) \oplus \Theta(f)$ is defined implies $e \oplus f$ is defined. 
An \emph{isomorphism} is a monomorphism that maps $\ea_1$ onto all of $\ea_2$.

\subsection{Undefined pairs}

The essence of an effect algebra structure is the fact that $\oplus$ is not a binary operation but a partial binary operation.
We say that a pair $(e,f)$ is \emph{undefined} if $e \oplus f$ is not defined, and in this case we denote $e \diamond f$.
The condition for a sum to be defined can also be expressed by using the partial order of the effects. 
Namely, one can verify the following:
\begin{enumerate}
\item[(EA-f)] $e \oplus f$ is defined if and only if $e\leq f^\bot$\, .
\end{enumerate}

For later use, we observe some conditions when a pair of effects is undefined.

\begin{proposition}\label{prop:defined}
Assume that  $e\oplus f$ is defined and $e \neq f^\bot$. 
Then $e^\bot \diamond f^\bot$.
\end{proposition}

\begin{proof}
Since $e \oplus f$ is defined, $e\leq f^\bot$. As $e \neq f^\bot$, we have $e < f^\bot$.
Let us make a counter assumption that $e^\bot \oplus f^\bot$ is defined. That implies 
$$
e^\bot \oplus f^\bot > e^\bot \oplus e = \eone \, ,
$$
which is a contradiction. 
\end{proof}

\begin{proposition}\label{prop:notdefined}
Let $e$ and $f$ be self-complementary effects and $e\neq f$.
Then $e \diamond f$.
\end{proposition}

\begin{proof}
Suppose $e \oplus f$ is defined. 
Since $e\neq f$ and $e$ is self-complementary, we have $e\neq f^\bot$.
It follows from Prop. \ref{prop:defined} that $e^\bot \diamond f^\bot$.
But $e^\bot =e$ and $f^\bot=f$, hence this is a contradiction.
\end{proof}

\begin{proposition}\label{prop:notself}
Assume $e \oplus e$ is defined and $e$ is not self-complementary.
Then $e^\bot \diamond e^\bot$. 
\end{proposition}

\begin{proof}
Since $e \oplus e$ is defined, $e \leq e^\bot$ by (EA-f).
Suppose $e^\bot \oplus e^\bot$ is defined.
Then $e^\bot \leq e$, hence $e^\bot = e$.
This contradicts $e$ not being self-complementary, thus $e^\bot \diamond e^\bot$. 
\end{proof}

\begin{proposition}\label{prop:atmostone}
There is at most one nontrivial effect $e$ such that $e \oplus f$ is defined for all $f\neq \eone$.
\end{proposition}

\begin{proof}
Let $e$ be a nontrivial effect such that the sum $e\oplus f$ is defined with all $f\neq \eone$.
Let $g\neq e$ be another effect with the same property. Then $g \oplus e^\bot$ is defined and $g \oplus e^\bot \neq \eone$ as the complement effect is unique.
Hence, $(g \oplus e^\bot) \oplus e$ is defined.
Then 
$$
(g \oplus e^\bot) \oplus e = g \oplus (e^\bot \oplus e) = g \oplus \eone
$$
is defined, which implies that $g=\enul$.

\end{proof}

Finally, there is a simple condition that guarantees that $e \oplus f$ is defined. 

\begin{proposition}\label{prop:eef}
Assume $f=e \oplus e$ and $f \oplus f$ is defined.
Then $e \oplus f$ is defined. 
\end{proposition}

\begin{proof}
Since $f \oplus f=(e \oplus e) \oplus f$  is defined, also $e \oplus (e \oplus f)$ is defined.
\end{proof}

\subsection{Effect subalgebras}

Loosely speaking, a subset of an effect algebra is a subalgebra if it is an effect algebra in a consistent way.
The fact that the effect sum is only partially defined gives two natural formulations of this concept.

\begin{definition}\label{def:subalgebra}
Let $(\ea,\enul,\eone,\oplus)$ be an effect algebra and $\ea' \subseteq \ea$.
We say that $(\ea',\enul,\eone,\oplus')$ is
\begin{itemize}
\item[(a)] an \emph{effect subalgebra} of $(\ea,\enul,\eone,\oplus)$ if $e \oplus' f$ is defined if and only if $e \oplus f$ is defined, in which case $e \oplus' f = e \oplus f$.
\item[(b)] a \emph{weak effect subalgebra} of $(\ea,\enul,\eone,\oplus)$ if $e \oplus' f$ is defined only if $e \oplus f$ is defined, in which case  $e \oplus' f = e \oplus f$.
\end{itemize}
\end{definition}
 
 The difference of these two concepts is that in a weak effect subalgebra some pairs can be undefined that give a defined sum in the original effect algebra.
 It is clear that an effect subalgebra of $\ea$ is a weak effect subalgebra, but the converse need not hold; we will demonstrate that shortly.
 Their difference becomes visible also in their intrinsic partial orderings, as shown next.  
 
 \begin{proposition}\label{prop:order}
 Let $\ea$ be an effect algebra and $\ea' \subseteq \ea$.
 \begin{itemize}
 \item[(a)]  If $\ea'$ is a weak effect subalgebra of $\ea$, then
 \begin{equation}
 e \leq' f \quad \Rightarrow \quad  e \leq f 
 \end{equation}
 for all $e,f \in \ea'$. 
 \item[(b)] If $\ea'$ is an effect subalgebra of $\ea$, then
 \begin{equation}
 e \leq' f \quad \Leftrightarrow \quad  e \leq f 
 \end{equation}
 for all $e,f \in \ea'$.
 \end{itemize}
 \end{proposition}
 
 \begin{proof}
 \begin{itemize}
 \item[(a)] Suppose $e \leq' f$. Then there exists $g\in \ea'$ such that $e\oplus' g = f$. 
 It follows that $e\oplus g$ is defined and $e\oplus g=f$.
 \item[(b)] Suppose $e \leq f$. Then $e \oplus f^\bot$ is defined. Since $f^\bot\in\ea'$, it follows that $e \oplus' f^\bot$ is defined.
 Hence, $e \leq' f$.
 \end{itemize}
 \end{proof}
 
 In Definition \ref{def:subalgebra} we have introduced two notions: effect subalgebra and weak effect subalgebra. In the following we apply these to the concrete effect algebras - the standard scale effect algebra $[0,1]$, the standard fuzzy set effect algebra $[0,1]^n$ and the standard quantum effect algebra $\eh$.
 
 \begin{definition}
 An effect algebra $\ea$ is 
 \begin{itemize}
 \item a \emph{(weak) scale effect algebra} if it is isomorphic to a (weak) effect subalgebra of $[0,1]$.
 \item a \emph{(weak) fuzzy set effect algebra} of dimension $n$ if it is isomorphic to a (weak) effect subalgebra of $[0,1]^n$.
 \item a \emph{(weak) quantum effect algebra} of dimension $d$ if it is isomorphic to a (weak) effect subalgebra of $\eh$ with $\dim\hi=d$.
 \end{itemize}
 \end{definition}
 
 In the following, when we say that a finite subset $X \subset [0,1]$ is an effect algebra, we mean that the effect sum of two elements $x,y \in X$ is defined if and only if $x+ y \in X$, and in that case it is the usual addition. 
 In order for $X$ to be an effect algebra, one has to check that associativity holds.
 An effect algebra of this form may be a scale effect algebra, or just a weak scale effect algebra.
 
\begin{example}
Examples of scale effect algebras are $\{0,\tfrac{1}{2},1\}$, $\{0,\tfrac{1}{3},\tfrac{2}{3},1\}$ and $\{0,\tfrac{1}{4},\tfrac{1}{2}, \tfrac{3}{4},1\}$.
Examples of weak scale effect algebras that are not scale effect algebras are $\{0,\tfrac{1}{5},\tfrac{1}{2}, \tfrac{4}{5},1\}$ and $\{0,\tfrac{1}{4},\tfrac{1}{3}, \tfrac{2}{3},\tfrac{3}{4},1\}$.
\end{example}

\begin{example}
Necessary conditions that a finite subset $X$ of $[0,1]$ is a weak scale effect algebra are that $0,1 \in X$ and $1-x \in X$ whenever $x \in X$. 
However, these conditions are not sufficient. For instance, let $X=\{0,\tfrac{1}{6},\tfrac{1}{3}, \tfrac{2}{3},\tfrac{5}{6}, 1 \}$. 
Then $(\tfrac{1}{6} \oplus \tfrac{1}{6}) \oplus \tfrac{1}{3} = \tfrac{2}{3}$, but $\tfrac{1}{6} \oplus \tfrac{1}{3}$ is not defined in $X$, therefore (EA2) does not hold.
\end{example}

\begin{proposition}\label{prop:anyis}
\begin{itemize}
\item[(a)] Any (weak) scale effect algebra is a (weak) fuzzy set effect algebra.
\item[(b)] Any (weak) fuzzy set effect algebra is a (weak) quantum effect algebra.
\end{itemize}
\end{proposition}

\begin{proof}
\begin{itemize}
\item[(a)] This is trivial as we can have $n=1$ in the definition of (weak) fuzzy set effect algebra.
\item[(b)] The isomorphism is given by the map that maps vectors $\vx \in \real^n$ into diagonal matrices $diag(x_1,\ldots,x_n)$. 
\end{itemize}
\end{proof}
We will see in Sec. \ref{sec:5elem}, a weak scale effect algebra is not necessarily a quantum effect algebra.

Finally, we make a remark on weak quantum effect algebras.
In the standard quantum effect algebra $\eh$, there is only one effect $E$ satisfying $E+E = \id$, namely, $E=\half \id$. 
Similarly, if two effects $E$ and $F$ satisfy $E+E=F+F$, then $E=F$.
Therefore, in a weak quantum effect algebra there can be at most one self-complementary element, and two elements must be the same if their multiples are the same.
We formulate this observation as a proposition for later use.

\begin{proposition}\label{prop:not-quantum}
Let $\ea$ be an effect algebra and $e,f\in\ea$ two different elements satisfying
\begin{equation*}
\overbrace{e \oplus e \oplus \cdots \oplus e}^n = \overbrace{f \oplus f \oplus \cdots \oplus f}^{n}
\end{equation*}
for some $n\geq 2$. Then $\ea$ is not a weak quantum effect algebra.
\end{proposition}

\section{Sum table}

We are going to concentrate on finite effect algebras, i.e., those where $\ea$ is a finite set.
When $\ea$ is small, it is illustrative to draw a table and mark there all effect sums.
Since $e \oplus \enul = e$ for all effects and $e \oplus \eone$ is defined only when $e=\enul$, these two rows and columns can be left out in the table.
For instance, suppose $\ea=\{ \enul,\eone,e,f,g \}$. 
The form of the table is then 
\begin{center}
\begin{tabular}{|c!{\boldv}c|c|c|  } 
\hline
\cplus $\oplus$ & \ccb{e}  & \ccy{f} & \ccr{g}\\
\boldh
 \ccb{e} & $\cdot$ & $\cdot$ & $\cdot$   \\
\hline
 \ccy{f} & $\cdot$ & $\cdot$  & $\cdot$   \\
\hline
 \ccr{g} & $\cdot$ & $\cdot$  & $\cdot$   \\
\hline
\end{tabular}
\end{center}
where the respective sums are marked in the places of $\cdot$.
When the sum is not defined, we mark it $\diamond$.
If the sum is defined but not specified, we mark it $\bullet$.
We call this table, when filled, the \emph{sum table} of $\ea$.
A possible sum table is, for instance, the following:
\begin{center}
\begin{tabular}{|c!{\boldv}c|c|c|  } 
\hline
\cplus $\oplus$ & \ccb{e}  & \ccy{f} & \ccr{g} \\
\boldh
\ccb{e} & \ccr{g} & \ccnot & \ccone   \\
\hline
\ccy{f} & \ccnot & \ccone  & \ccnot   \\
\hline
\ccr{g} & \ccone & \ccnot  & \ccnot   \\
\hline
\end{tabular}
\end{center}
The sum table specifies all the sums, hence it uniquely identifies $\ea$, up to isomorphism.
One can confirm that this table, indeed, defines a valid effect algebra. We will come later to ways to check that a given table corresponds to an actual effect algebra.

There are some simple rules that follow from the effect algebra axioms and derived properties:
\begin{itemize}
\item[(ET1)] table is symmetric (because of (EA1))
\item[(ET2)] at each row there is $\eone$ in exactly one position (because of (EA3))
\item[(ET3)] $\enul$ does not appear in the table (because of (EA-d))
\item[(ET4)] $e$ is not in the row or in the column labeled with $e$ (because of (EA-c))
\end{itemize}

We also note that Prop. \ref{prop:atmostone} means that there can at most one row without any $\diamond$ symbols.
Further, from Prop. \ref{prop:notdefined} we obtain the following rule for filling of sum tables: 
\begin{itemize}
\item[(ET5)]
\qquad
\begin{tabular}{|c|c|  } 
\hline
$\eone$ & $\cdot$ \\
\hline
 $\cdot$ & $\eone$   \\
\hline
\end{tabular}
\quad$\Rightarrow$\quad
\begin{tabular}{|c|c|  } 
\hline
$\eone$ & $\diamond$ \\
\hline
 $\diamond$ & $\eone$   \\
\hline
\end{tabular}
\end{itemize}
In a similar way, Prop. \ref{prop:notself} gives the following rule:
\begin{itemize}
\item[(ET6)]
\qquad
\begin{tabular}{|c|c|  } 
\hline
$\bullet$ & $\eone$  \\
\hline
 $\eone$ & $\cdot$   \\
\hline
\end{tabular}
\quad$\Rightarrow$\quad
\begin{tabular}{|c|c|  } 
\hline
$\bullet$ & $\eone$  \\
\hline
 $\eone$ & $\diamond$   \\
\hline
\end{tabular}
\end{itemize}
From Prop. \ref{prop:eef} we conclude the following rule:
\begin{itemize}
\item[(ET7)]
\qquad
\begin{tabular}{|c!{\boldv}c|c|c|  } 
\hline
\cplus $\oplus$ & e  & f \\
\boldh
 e & f & $\cdot$  \\
\hline
 f & $\cdot$ & $\bullet$   \\
 \hline
\end{tabular}
\quad$\Rightarrow$\quad
\begin{tabular}{|c!{\boldv}c|c|c|  } 
\hline
\cplus $\oplus$ & e  & f \\
\boldh
 e & f & $\bullet$  \\
\hline
 f & $\bullet$ & $\bullet$   \\
 \hline
\end{tabular}

\end{itemize}

Not any filling satisfying the above rules determines a valid effect algebra. 
The reason is that the associativity of $\oplus$ is not visible from the sum table and must be confirmed separately.
By a \emph{model} for a sum table we mean a concrete effect algebra that consists of some concrete objects (e.g. numbers, vectors, matrices) and where $\oplus$ is a restriction of a commutative and associative operation.
By providing a model is a tool to verify that the effect algebra axioms are satisfied and that a table determines a valid sum table.
By a \emph{real model} or \emph{complex model} we mean a finite subset $Z\subset \real$ or $Z \subset \complex$, such that the effect sum is the ordinary multiplication and the effect sum $z\oplus w$ is defined if and only if $zw \in Z$. The zero effect is $1$ while the unit effect can vary, depending on the model.
In a similar way, by a \emph{vector model} we mean a finite subset of either $\real^n$ or $\complex^n$, where the entrywise multiplication is the effect sum.
We will later provide models for sum tables.

Two different sum tables may represent two isomorphic effect algebras. 
For instance, the sum table
\begin{center}
\begin{tabular}{|c!{\boldv}c|c|c|  } 
\hline
\cplus $\oplus$ & \ccb{e}  & \ccy{f} & \ccr{g} \\
\boldh
\ccb{e} & \ccone & \ccnot & \ccnot   \\
\hline
\ccy{f} & \ccnot & \ccr{g}  & \ccone   \\
\hline
\ccr{g} & \ccnot & \ccone  & \ccnot   \\
\hline
\end{tabular}
\end{center}
is obtained from the previously depicted sum table by relabeling $e$ and $f$, hence it describes the same effect algebra structure.
Generally, by a \emph{switch} for a sum table we mean an interchange of a row $i$ with row $j$ and an interchange of column $i$ with column $j$.
This corresponds to a relabelling of two effects.
Two sum tables $T_1$ and $T_2$ represent isomorphic effect algebras if and only if $T_1$ can be obtained from $T_2$ by a sequence of switches.

\section{Finite scale effect algebras}

We recall that a scale effect algebra is an effect subalgebra of the standard scale effect algebra $[0,1]$.
Let $n \geq 2$.
We denote by $\sea{n}$ the scale effect algebra consisting of the following $n$ elements: $\{ 0, \tfrac{1}{n-1} , \tfrac{2}{n-1}, \ldots,  \tfrac{n-2}{n-1}, 1\}$.

\begin{theorem}\label{thm:scale}
Let $\ea$ be an effect algebra with $n$ elements.
The following statements are equivalent.
\begin{itemize}
\item[(i)] $\ea$ is a scale effect algebra.
\item[(ii)] $\ea$ is totally ordered.
\item[(iii)] $\ea$ is generated by a single element, i.e., $\ea$ consists only of sums $e\oplus \cdots \oplus e$ for a fixed element $e$.
\item[(iv)] $\ea$ is isomorphic to $\sea{n}$.
\end{itemize}
In particular, for every $n \geq 2$ there is a unique scale effect algebra with $n$ elements.
\end{theorem}

\begin{proof}
(i)$\Rightarrow$(ii): A scale effect algebra with $n$ elements is of the form $\ea=\{ 0, x_1,x_2,\ldots, x_{n-2},1\} \subset \real$. 
Without loss of generality, we can assume that $0<x_1<x_2< \cdots <x_{n-2}<1$.
Hence, $\ea$ is totally ordered.
\newline
(ii)$\Rightarrow$(iii): A totally ordered effect algebra $\ea$ with $n$ elements is of the form $\ea=\{ \enul, e_1,e_2,\ldots, e_{n-2},\eone\}$ with $\enul<e_1<e_2< \cdots <e_{n-2}<\eone$.
Since $e_1 < e_2$, we have that $e_2=e_1 \oplus e_i$ for some $i \in \{1,2,\ldots,n-2 \}$.
Since $e_i < e_i \oplus e_1 =e_2$, we conclude that $e_i=e_1$ since this is the only nonzero element below $e_2$.
It follows that $e_1 \oplus e_1 = e_2$.
Continuing, we have that $e_2 < e_3$, hence $e_3=e_2 \oplus e_j$ for some $j \in \{1,2,\ldots,n-2 \}$.
As $e_j < e_j \oplus e_2 = e_3$, we conclude that $e_j = e_1$ or $e_j = e_2$.
Let us first assume that $e_j=e_2$. Then $e_3=e_2 \oplus e_2 = (e_1 \oplus e_1) \oplus e_2$.
It follows that $e_1 \oplus e_2$ is defined. 
Then
$$
e_1 < e_2 < e_1 \oplus e_2 < e_2 \oplus e_2 = e_3 \, ,
$$
which contradicts the fact that there is no element between $e_2$ and $e_3$. Hence, $e_j=e_1$.
This means that $e_3 = e_1 \oplus e_1 \oplus e_1$. 
We can continue by induction to prove that $e_\ell = e_1 \oplus \cdots \oplus e_1$, where there are $\ell$ $e_1$'s.
Therefore, $\ea$ is generated by $e_1$.
(iii)$\Rightarrow$(iv): Suppose $\ea$ is generated by a single element $e$.
Any other element is then a sum $\ell e \equiv e \oplus \cdots \oplus e$ for some $\ell \geq 2$.
By associativity, if $\ell e$ is defined, then also $(\ell-1)e$ is defined. 
The map $\Phi:\ea \to \sea{n}$ defined as $\Phi(\ell e) = \tfrac{\ell}{n-1}$ is an isomorphism.
\newline
(iv)$\Rightarrow$(i): Trivial.
\end{proof}

By examining the sum table of $\sea{n}$, we observe that the anti-diagonal consists of $\eone$, the upper left corner is filled with nontrivial effects whereas the lower right corner is filled with $\diamond$.
For instance, the sum table of $\sea{6}$ is 
\begin{center}
\begin{tabular}{|c!{\boldv} c|c|c| c|   } 
\hline
\cplus $\oplus$ & \ccb{e}  & \ccg{f} & \ccy{g}  & \ccr{h} \\
\boldh
\ccb{e} & \ccg{f} & \ccy{g} & \ccr{h} & \ccone   \\
\hline
\ccg{f} & \ccy{g}  & \ccr{h} &  \ccone & \ccnot   \\
\hline
\ccy{g} & \ccr{h} & \ccone & \ccnot  & \ccnot    \\
\hline
\ccr{h}  & \ccone & \ccnot  & \ccnot & \ccnot \\
\hline
\end{tabular}
\end{center}
The fact that one row is full (i.e. no $\diamond$ symbols) is a characteristic feature of the sum tables of scale effect algebras.

\begin{proposition}
A sum table corresponds to a scale effect algebra if and only if there is one row without $\diamond$ symbols.
\end{proposition}

\begin{proof}
By Theorem \ref{thm:scale}, a scale effect algebra $\sea{n}$ is generated by a single element $e$.
All other $n-2$ nontrivial effects are of the form $e\oplus \cdots \oplus e$, hence the sum with $e$ is defined.
This means that there is one row without $\diamond$ symbols.

Suppose that we have a sum table and one row has no $\diamond$ symbols.
Let $e$ be the effect that labels this row, which means that $e \oplus f$ is defined for all $f \neq \eone$.
In particular, $e \oplus e$, $e \oplus e \oplus e$, .., and finally $ne \equiv e \oplus \cdots \oplus e=\eone$ are defined. 
If $e \oplus f \neq \eone$, then $e \oplus (e \oplus f)$ is defined.
Eventually, $e \oplus \cdots \oplus e \oplus f = \eone$ for some number, say $k$, of $e$ effects.
It follows that $f=(n-k)e$.
This shows that the effect algebra is generated by $e$, hence by Theorem \ref{thm:scale} it is a scale effect algebra.
\end{proof}

Let us then focus on the number of defined sums.
In $\sea{n}$, among the $(n-2)^2$ nontrivial pairs (i.e. those where both elements are nontrivial) there are $\half(n-1)(n-2)$ defined sums and $\half(n-2)(n-3)$ undefined pairs.
In the following we show that among all effect algebras with $n$ elements, $\sea{n}$ has the maximal number of pairs that are defined and this is a characteristic feature of $\sea{n}$.

\begin{theorem}\label{thm:atmost}
An effect algebra $\ea$ with $n$ elements has at most ${\half(n-1)(n-2)}$ nontrivial defined sums.
If it has $\half(n-1)(n-2)$ nontrivial defined sums, then it is the scale effect algebra $\sea{n}$.
\end{theorem}

\begin{proof}
We write $\ea$ as $\ea=\{\enul,\eone,e_1,e_1^\bot,\ldots,e_r,e_r^\bot,f_1,\ldots,f_s \}$, where $f_1,\ldots,f_s$ are self-complementary.
To calculate the number of nontrivial undefined pairs, we divide all nontrivial pairs into five cases:
\begin{itemize}
\item[(1)] Among the pairs $(e_i,e_j)$ and $(e_i^\bot,e_j^\bot)$, at least $r^2$ are undefined by Prop. \ref{prop:defined}.
\item[(2)] Among the pairs $(e_i,e_j^\bot)$ and $(e_i^\bot,e_j)$, $i \neq j$, at least $r(r-1)$ are undefined by Prop. \ref{prop:defined}.
\item[(3)] Among the pairs  $(e_i,f_j)$ and $(e_i^\bot,f_j)$, at least $rs$ are undefined by Prop. \ref{prop:defined}.
\item[(4)] Among the pairs  $(f_j,e_i)$ and $(f_j,e_i^\bot)$, at least $rs$ are undefined by Prop. \ref{prop:defined}.
\item[(5)] All the pairs $(f_i,f_j)$, $i \neq j$, are undefined by Prop. \ref{prop:notdefined}. Hence, this gives exactly $s(s-1)$ undefined pairs.
\end{itemize}
Adding these five cases together we conclude that there are at least $u \equiv (r+s)^2+r(r-1)-s$ undefined nontrivial pairs.
Since $2r+s=n-2$, we have
\begin{align*}
u - \half (n-2)(n-3) &= (r+s)^2+r(r-1)-s - (2r+s)(2r+s-1)/2 \\
& = \half s(s-1) \geq 0 \, .
\end{align*}
Therefore, there are at least $\half (n-2)(n-3)$ undefined nontrivial pairs, which means that there are at least $\half(n-1)(n-2)$ nontrivial defined sums.

To prove the second statement, we observe that to have the minimal number of undefined pairs we must have $s=0$ or $s=1$. Further, we must have exactly $r^2$ undefined pairs in the case (1) above, which means that either $e_i \oplus e_j$ or $e_i^\bot \oplus e_j^\bot$ is defined.
Therefore, either $e_i \leq e_j^\bot$ or $e_j^\bot \leq e_i$.
In a similar way we must have exactly $r(r-1)$ undefined pairs in the case (2) above, which implies either $e_i \oplus e_j^\bot$ or $e_i^\bot \oplus e_j$ is defined, hence $e_i \leq e_j$ or $e_j \leq e_i$.
Finally, we must have exactly $rs$ undefined pairs in the case (3) above, which implies either $e_i \oplus f_j$ or $e_i^\bot \oplus f_j$ is defined, hence $e_i \leq f_j$ or $f_j \leq e_i$.
We conclude that $\ea$ is totally ordered. 
From Theorem \ref{thm:scale} follows that $\ea$ is a scale effect algebra.
\end{proof}

Finally, for later use we observe a simple condition that may identify when an effect algebra is not a scale effect algebra already from two elements and their relation.

\begin{proposition}\label{prop:not-scale}
Suppose that an effect algebra $\ea$ contains two elements $e$ and $f$ such that $e \oplus f = \eone$ while $e \diamond e$ and $f \diamond f$. Then $\ea$ is not a scale effect algebra.
\end{proposition}

\begin{proof}
Let us consider a scale effect algebra $\sea{n}$ with two elements $x,y \in \sea{n}$ such that $x+ y = 1$.
We can assume that $x \leq y $, hence it follows that $x \leq \tfrac{1}{2}$.
Therefore, $x + x \in [0,1]$, which implies that $x + x \in \sea{n}$.
 \end{proof}

\section{Sparse finite effect algebras}

Every effect has a complement effect and therefore $\eone$ appears in every row of a sum table.
We say that an effect algebra $\ea$ is \emph{sparse} if, apart from complements and addition with $\enul$, no other sums are defined. 
This means that among $(n-2)^2$ nontrivial pairs, only $n-2$ are defined.
In the following we first introduce certain separate classes of sparse effect algebras, the first having diagonal sum table and the second one consisting of complementary pairs.
The general sparse effect algebra is then a combination of these two.

\subsection{Sparse effect algebra $\dea{n}$}\label{sec:sparse1}

Let $n \geq 3$ and suppose there are $n$ elements $\{ \enul,\eone,e_1,\ldots,e_{n-2} \}$ such that each $e_j$ is self-complementary, i.e.,  $e_j^\bot=e_j$ for all $j=1,\ldots,n-2$. 
It follows from Prop. \ref{prop:notdefined} that there are no other nontrivial sums defined, meaning that $e_j \diamond e_k$ for all $j\neq k$. 
We denote by $\dea{n}$ this effect algebra.
The sum table of $\dea{n}$ contains $\eone$ in the diagonal and $\diamond$ elsewhere.
For instance, the sum table of $\dea{5}$ is
\begin{center}
\begin{tabular}{|c!{\boldv}c|c|c|  } 
\hline
\cplus $\oplus$ & \ccb{e}  & \ccy{f} & \ccr{g} \\
\boldh
\ccb{e} & \ccone & \ccnot & \ccnot   \\
\hline
\ccy{f} & \ccnot & \ccone  & \ccnot   \\
\hline
\ccr{g} & \ccnot & \ccnot  & \ccone   \\
\hline
\end{tabular}
\end{center}

To see that $\dea{n}$ is a valid effect algebra, we write a vector model for it.
We choose $\dea{n}$ to be a subset of vectors in $\real^{n-2}$ and $\oplus$ the componentwise multiplication. The effect $\enul$ corresponds to the vector $(1,\ldots,1)$ and $\eone$ to the vector $(4,\ldots,4)$. 
The effect $e_j$ corresponds to the vector that has $-2$ in the $j$th entry and $2$ in all other entries, and there are no other elements.
For instance, for $\dea{5}$ this means that
$$
e_1  \leftrightarrow (-2, 2,2), \quad e_2  \leftrightarrow (2,-2,2), \quad e_3  \leftrightarrow (2,2,-2) \, .
$$
It is clear that $e_j \oplus e_j = (4,\ldots,4)$ while the componentwise multiplications do not give a valid element.
Hence, this is a model for $\dea{n}$.

The effect algebra $\dea{3}$ is the scale effect algebra $\{0,\half,1\}$.
The effect algebra $\dea{n}$ with $n \geq 4$ is not a quantum effect algebra, and not even a weak quantum effect algebra.
This follows from Prop. \ref{prop:not-quantum} as there are (at least) two different effects $e_1$ and $e_2$ that satisfy $e_1\oplus e_1=e_2 \oplus e_2 = \eone$. 

\subsection{Sparse effect algebra $\pea{n}$}\label{sec:non-inv}

Suppose $\ea=\{ \enul,\eone,e_1,e_1^\bot,\ldots, e_\ell,e_\ell^\bot \}$ such that $e_j \diamond e_k$ for all $j,k$ and  $e_j \diamond e^\bot_k$ for all $j\neq k$.
We denote by $\pea{n}$ this effect algebra with $\ell$ complementary pairs and hence with $n=2\ell +2$ elements.
For instance, the sum table of $\pea{6}$ is
\begin{center}
\begin{tabular}{ |c!{\boldv} c|c|c| c|   } 
\hline
\cplus $\oplus$ & \ccb{e}  & \ccg{f} & \ccy{g}  & \ccr{h} \\
\boldh
\ccb{e} & \ccnot & \ccone & \ccnot & \ccnot   \\
\hline
\ccg{f} & \ccone & \ccnot & \ccnot & \ccnot    \\
\hline
\ccy{g} & \ccnot & \ccnot & \ccnot  & \ccone   \\
\hline
\ccr{h} & \ccnot & \ccnot   & \ccone & \ccnot  \\
\hline
\end{tabular}
\end{center}

The effect algebra $\pea{n}$ is not a scale effect algebra (e.g. by Prop. \ref{prop:not-scale}).
However, it is a fuzzy set effect algebra of dimension 2 with
\begin{align*}
\vec{e}_1 &= (\tfrac{1}{3},\tfrac{2}{3}) \, ,  &\vec{e}_1^\bot &= (\tfrac{2}{3},\tfrac{1}{3})  \\
\vec{e}_2 &= (\tfrac{1}{4},\tfrac{3}{4}) \, ,  &\vec{e}_2^\bot &= (\tfrac{3}{4},\tfrac{1}{4})  \\
& \vdots & \vdots \\
\vec{e}_\ell &= (\tfrac{1}{\ell+2},\tfrac{\ell+1}{\ell+2}) \, ,  &\vec{e}_\ell^\bot &= (\tfrac{\ell+1}{\ell+2},\tfrac{1}{\ell+2}) \, .
\end{align*}

\subsection{Sparse effect algebra $\emix{n(k,\ell)}$}

In general, a sparse effect algebra is a combination of the previous two structures.
In a sparse effect algebra each row in the sum table has $\eone$ and no other effects.
Let $\ea=\{\enul,\eone,e_1,\ldots,e_k,f_1,f_1^\bot,\ldots,f_\ell,f_\ell^\bot \}$ and suppose that $\{\enul,\eone,e_1,\ldots,e_k \}$ is the sparse effect algebra $\dea{k+2}$ and $\{\enul,\eone,f_1,f_1^\bot,\ldots,f_\ell,f_\ell^\bot \}$ is the sparse effect algebra $\pea{2\ell+2}$.
We denote the total effect algebra $\ea$ as $\emix{n(k,\ell)}$.

For instance, the sum table of $\emix{5(1,1)}$ is
\begin{center}
\begin{tabular}{|c!{\boldv}c|c|c|  } 
\hline
\cplus $\oplus$ & \ccb{e}  & \ccy{f} & \ccr{g} \\
\boldh
\ccb{e} & \ccone & \ccnot & \ccnot   \\
\hline
\ccy{f} & \ccnot & \ccnot & \ccone    \\
\hline
\ccr{g} & \ccnot & \ccone & \ccnot     \\
\hline
\end{tabular}
\end{center}

A vector model for $\emix{n(k,\ell)}$ is the following. 
We choose $\ea$ to be a subset of vectors in $\real^{k+\ell}$, $\oplus$ is the componentwise multiplication, $\enul$ corresponds to the vector $(1,\ldots,1)$ and $\eone$ to the vector $(4,\ldots,4)$. 
An effect $e_j$ corresponds to the vector that has $-2$ in the $j$th entry and $2$ in other entries. 
An effect $f_j$ corresponds to the vector that has $-2i $ in the $(k+j)$th entry and $2$ in other entries, while $f_{j}^\bot$ corresponds to the vector that has $2i $ in the $(k+j)$th entry and $2$ in other entries.
For instance, in $\emix{6(2,1)}$ we have
$$
e_1  \leftrightarrow (-2, 2,2) \, , \quad e_2  \leftrightarrow (2,-2,2) \, ,\quad  f_1  \leftrightarrow (2, 2,-2i), \quad f_2  \leftrightarrow (2,2,2i) \, .
$$

\begin{proposition}
The number of sparse effect algebras (up to isomorphism) with $n$ elements is 
\begin{align*}
\textrm{$\tfrac{n}{2}$ if $n$ is even.} \\
\textrm{$\tfrac{n-1}{2}$ if $n$ is odd.}
\end{align*}
\end{proposition}

\begin{proof}
The general sparse effect algebra with $n$ elements is $\emix{n(k,\ell)}=\{\enul,\eone,e_1,\ldots,e_k,f_1,f_1^\bot,\ldots,f_\ell,f_\ell^\bot \}$, where $n=2+k+2\ell$.
We note that $\ell$ is determined if $n$ and $k$ are fixed.

Suppose $n$ is even. Then $k$ is even and has possible values $0,2,4,\ldots,n-2$.
Hence, there are $\tfrac{n}{2}$ sparse effect algebras.

Suppose $n$ is odd. Then $k$ is odd and has possible values $1,3,5,\ldots,n-2$.
Hence, there are $\tfrac{n-1}{2}$ sparse effect algebras.
\end{proof}

\begin{example}
There are in total 4 sparse effect algebras with 8 elements. 
These are $\emix{8(6,0)}=\dea{8}$, $\emix{8(4,1)}$, $\emix{8(2,2)}$ and $\emix{8(0,4)}=\pea{8}$.
\end{example}

\section{Classification of small effect algebras}\label{sec:small}

In this section we list all effect algebras with 2--6 elements by giving their effect sum tables.
We classify them into those that are quantum effect algebras and those that are not.
We also present some examples of tables that might look natural but are not valid effect sum tables. 
The effect algebras with 2--6 elements are listed in Table \ref{table2}. 
We recall that $\sea{n}$ denotes the scale effect algebra with $n$ elements.
The sparse effect algebras come in three different families and they are denoted by $\dea{n}$, $\pea{n}$ and $\emix{n(k,\ell)}$.
Those effect algebras that do not belong to these four families or are not their compositions are listed as $\ea_{n(m)}$, where $n$ is the number of elements and $m$ the index in the following list.
The compositional structure is explained in Sec. \ref{sec:composition}.

\begin{table}[htbp]
\centering
\caption{Summary of effect algebras with 2--6 elements.}
\label{table2}
\begin{tabular}{|c|c|c|}
\hline
\textbf{\# elements} &  \textbf{quantum} & \textbf{non-quantum} \\ 
\boldh
2  & $\sea{2}$ & -  \\ \hline
3  & $\sea{3}$ & -  \\ \hline
4  &  $\sea{4}$, $\pea{4}=\sea{2}\times\sea{2}$ & $\dea{4}$  \\ \hline
5  & $\sea{5}$, $\emix{5(1,1)}$  & $\dea{5}$, $\ea_{5(3)}$ \\ \hline
6  & $\sea{6}$, $\sea{2}\times\sea{3}$, $\pea{6}$, $\ea_{6(4)}$ & $\dea{6}$, $\emix{6(2,1)}$, $\ea_{6(5)}$, $\ea_{6(6)}$, $\ea_{6(8)}$, $\ea_{6(9)}$  \\ \hline
\end{tabular}
\end{table}

\subsection{Two elements} The unique two element effect algebra is the scale effect algebra $\scale{2}=\{0,1\}$.

\subsection{Three elements}  Let $\ea=\{ \enul,\eone,e \}$. The effect $e$ must have a complement effect and there are no other options than $e^\bot = e$. Hence, there is only one effect algebra and this is $\dea{3}$, which is the same as the scale effect algebra $\scale{3}=\{0,\half,1 \}$.

\subsection{Four elements} Let $\ea=\{ \enul,\eone,e,f \}$. 
Following the rules (ET1)--(ET6) we obtain three sum tables, up to permutation of the elements.

\begin{enumerate}

\item[(4-1)] The first option is 

\begin{center}
\begin{tabular}{|c!{\boldv}c|c| } 
\hline
\cplus $\oplus$ & \ccb{e}  & \ccr{f} \\
\boldh
\ccb{e} & \ccone & \ccnot   \\
\hline
\ccr{f} & \ccnot & \ccone    \\
\hline
\end{tabular}
\end{center}
This is the sparse effect algebra $\dea{4}$. It is not a weak quantum effect algebra as explained in Sec. \ref{sec:sparse1}.

\item[(4-2)] The second option is

\begin{center}
\begin{tabular}{|c!{\boldv}c|c| } 
\hline
\cplus $\oplus$ & \ccb{e}  & \ccr{f} \\
\boldh
\ccb{e} & \ccnot & \ccone   \\
\hline
\ccr{f} & \ccone & \ccnot    \\
\hline
\end{tabular}
\end{center}
This is the sparse effect algebra $\pea{4}$.
It is a fuzzy set effect algebra of dimension $2$ with $\vec{e} = (1,0)$ and $\vec{f} = (0,1)$. 

\item[(4-3)] The third option is

\begin{center}
\begin{tabular}{|c!{\boldv}c|c| } 
\hline
\cplus $\oplus$ & \ccb{e}  & \ccr{f} \\
\boldh
\ccb{e} & \ccr{f} & \ccone   \\
\hline
\ccr{f} & \ccone & \ccnot    \\
\hline
\end{tabular}
\end{center}
This is the scale effect algebra $\scale{3}=\{0,\half,1\}$.
\end{enumerate}

\noindent
The remaining way to fill the table is

\begin{enumerate}
\item[($\spadesuit$)] 
\begin{center}
\begin{tabular}{|c!{\boldv}c|c| } 
\hline
\cplus $\oplus$ & \ccb{e}  & \ccr{f} \\
\boldh
\ccb{e} & \ccr{f} & \ccone   \\
\hline
\ccr{f} & \ccone & \csum \ccb{e}    \\
\hline
\end{tabular}
\end{center}
\end{enumerate}
but this does not correspond to an effect algebra as it is in conflict with (ET6), or with Theorem \ref{thm:atmost}.
Alternatively, we can see that the associative law does not hold: $(e \oplus e) \oplus f$ is defined but $e \oplus (e \oplus f)$ is not.

\subsection{Five elements} \label{sec:5elem}
Let $\ea=\{ 0,1,e,f,g \}$. 
First of all, we have two sparse 5-element effect algebras: 
\begin{enumerate}
\item[(5-1)] $\dea{3}$, which has the sum table
\begin{center}
\begin{tabular}{|c!{\boldv}c|c|c|  } 
\hline
\cplus $\oplus$ & \ccb{e}  & \ccy{f} & \ccr{g} \\
\boldh
\ccb{e} & \ccone & \ccnot & \ccnot   \\
\hline
\ccy{f} & \ccnot & \ccone  & \ccnot   \\
\hline
\ccr{g} & \ccnot & \ccnot  & \ccone   \\
\hline
\end{tabular}
\end{center}
This is not a weak quantum effect algebra as explained in Sec. \ref{sec:sparse1}.

\item[(5-2)] $\emix{5(1,1)}$, which has the sum table
\begin{center}
\begin{tabular}{|c!{\boldv}c|c|c|  } 
\hline
\cplus $\oplus$ & \ccb{e}  & \ccy{f} & \ccr{g} \\
\boldh
\ccb{e} & \ccone & \ccnot & \ccnot   \\
\hline
\ccy{f} & \ccnot & \ccnot & \ccone    \\
\hline
\ccr{g} & \ccnot & \ccone & \ccnot     \\
\hline
\end{tabular}
\end{center}
This is a fuzzy set effect algebra with $\vec{e}=(\half,\half)$, $\vec{f}=(1,0)$, $\vec{g}=(0,1)$.
\end{enumerate}

In other effect algebras there are some nontrivial effects in the sum table.
From Theorem \ref{thm:atmost} we know that there are at least 3 undefined sums. By going through all possible ways to fill a table that are consistent with (ET1)--(ET6), we find two additional effect algebras.

\begin{enumerate}

\item[(5-3)] One option is

\begin{center}
\begin{tabular}{|c!{\boldv}c|c|c|  } 
\hline
\cplus $\oplus$ & \ccb{e}  & \ccy{f} & \ccr{g} \\
\boldh
\ccb{e} & \ccr{g} & \ccnot & \ccone   \\
\hline
\ccy{f} & \ccnot & \ccone  & \ccnot   \\
\hline
\ccr{g} & \ccone & \ccnot  & \ccnot   \\
\hline
\end{tabular}
\end{center}
This is a weak scale effect algebra $\{0,\tfrac{1}{3},\tfrac{1}{2},\tfrac{2}{3},1 \} \subset [0,1]$.
It is not a quantum effect algebra. 
To see this, assume it is. It follows from the sum table that $E=\tfrac{1}{3} \id$ and $F= \tfrac{1}{2} \id$, hence $E + F \leq \id$. 
But $E \oplus F$ is not defined, hence the assumption is absurd.

\item[(5-4)] One option is

\begin{center}
\begin{tabular}{|c!{\boldv}c|c|c|  } 
\hline
\cplus $\oplus$ & \ccb{e}  & \ccy{f} & \ccr{g} \\
\boldh
\ccb{e} & \ccy{f} & \ccr{g} & \ccone   \\
\hline
\ccy{f} & \ccr{g} & \ccone  & \ccnot   \\
\hline
\ccr{g} & \ccone & \ccnot  & \ccnot   \\
\hline
\end{tabular}
\end{center}
This is the scale effect algebra $\scale{5}=\{0,\tfrac{1}{4},\tfrac{1}{2},\tfrac{3}{4},1 \}$. 

\end{enumerate}

\noindent
For illustration, we list some other tables that are consistent with (ET1)--(ET4) but fail to be valid sum tables as they are in conflict with (ET5), (ET6) and (ET7), respectively.
\begin{enumerate}
\item[($\spadesuit$)] 
\begin{center}
\begin{tabular}{|c!{\boldv}c|c|c|  } 
\hline
\cplus $\oplus$ & \ccb{e}  & \ccy{f} & \ccr{g} \\
\boldh
\ccb{e} & \ccone & \ccr{g} & \ccnot   \\
\hline
\ccy{f} & \ccr{g} & \ccone  & \ccnot   \\
\hline
\ccr{g} & \ccnot & \ccnot  & \ccone   \\
\hline
\end{tabular}
\qquad
\begin{tabular}{|c!{\boldv}c|c|c|  } 
\hline
\cplus $\oplus$ & \ccb{e}  & \ccy{f} & \ccr{g} \\
\boldh
\ccb{e} & \ccnot & \ccr{g} & \ccone   \\
\hline
\ccy{f} & \ccr{g} & \ccone  & \ccnot   \\
\hline
\ccr{g} & \ccone & \ccnot  & \ccnot   \\
\hline
\end{tabular}
\qquad
\begin{tabular}{|c!{\boldv}c|c|c|  } 
\hline
\cplus $\oplus$ & \ccb{e}  & \ccy{f} & \ccr{g} \\
\boldh
\ccb{e} & \ccy{f} & \ccnot & \ccone   \\
\hline
\ccy{f} & \ccnot & \ccone  & \ccnot   \\
\hline
\ccr{g} & \ccone & \ccnot  & \ccnot   \\
\hline
\end{tabular}
\end{center}
\end{enumerate}
One can also see that these sum tables violate associativity.
For instance, in the first table $(e \oplus f) \oplus g$ is defined but $e \oplus (f \oplus g)$ is not.

\subsection{Six elements}

Let $\ea=\{ 0,1,e,f,g,h\}$. 
First of all, we have three sparse effect algebras with 6-elements:
\begin{enumerate}
\item[(6-1)] $\dea{6}$, which has the sum table
\begin{center}
\begin{tabular}{ |c!{\boldv} c|c|c| c|   } 
\hline
\cplus $\oplus$ & \ccb{e}  & \ccg{f} & \ccy{g}  & \ccr{h} \\
\boldh
\ccb{e} & \ccone & \ccnot & \ccnot  & \ccnot \\
\hline
\ccg{f} & \ccnot & \ccone  & \ccnot & \ccnot   \\
\hline
\ccy{g} & \ccnot & \ccnot  & \ccone & \ccnot  \\
\hline
\ccr{h} & \ccnot & \ccnot  &  \ccnot & \ccone   \\
\hline
\end{tabular}
\end{center}
This is not a weak quantum effect algebra as explained in Sec. \ref{sec:sparse1}.

\item[(6-2)] $\pea{6}$, which has the sum table
\begin{center}
\begin{tabular}{ |c!{\boldv} c|c|c| c|   } 
\hline
\cplus $\oplus$ & \ccb{e}  & \ccg{f} & \ccy{g}  & \ccr{h} \\
\boldh
\ccb{e} & \ccnot & \ccone & \ccnot & \ccnot   \\
\hline
\ccg{f} & \ccone & \ccnot & \ccnot & \ccnot    \\
\hline
\ccy{g} & \ccnot & \ccnot & \ccnot  & \ccone   \\
\hline
\ccr{h} & \ccnot & \ccnot   & \ccone & \ccnot  \\
\hline
\end{tabular}
\end{center}
This is a fuzzy set effect algebra as explained in Sec. \ref{sec:non-inv}.

\item[(6-3)] $\emix{6(2,1)}$, which has the sum table
\begin{center}
\begin{tabular}{ |c!{\boldv} c|c|c| c|   } 
\hline
\cplus $\oplus$ & \ccb{e}  & \ccg{f} & \ccy{g}  & \ccr{h} \\
\boldh
\ccb{e} & \ccone & \ccnot & \ccnot & \ccnot   \\
\hline
\ccg{f} & \ccnot & \ccone & \ccnot  & \ccnot  \\
\hline
\ccy{g} & \ccnot & \ccnot & \ccnot & \ccone   \\
\hline
\ccr{h} & \ccnot & \ccnot  & \ccone & \ccnot  \\
\hline
\end{tabular}
\end{center}
There are two self-complementary elements and it is therefore not a weak quantum effect algebra, as explained in Sec. \ref{sec:sparse1}.
\end{enumerate}

By going through the possible sum tables, one finds that there are 6 other effect algebras.
From Theorem \ref{thm:atmost} we know that there are at least 6 undefined sums.
 In the following we list the valid sum tables in the decreasing order of numbers of $\diamond$ in the table.

\begin{enumerate}

\item[(6-4)] One option is
\begin{center}
\begin{tabular}{|c!{\boldv} c|c|c| c|   } 
\hline
\cplus $\oplus$ & \ccb{e}  & \ccg{f} & \ccy{g}  & \ccr{h} \\
\boldh
\ccb{e} & \ccg{f} & \ccone & \ccnot & \ccnot   \\
\hline
 \ccg{f} & \ccone & \ccnot & \ccnot & \ccnot    \\
\hline
 \ccy{g} & \ccnot & \ccnot & \ccnot  & \ccone   \\
\hline
 \ccr{h} & \ccnot & \ccnot   & \ccone & \ccnot  \\
\hline
\end{tabular}
\end{center}
This is a fuzzy set effect algebra with $\vec{e}=(\tfrac{1}{3},\tfrac{1}{3})$, $\vec{f}=(\tfrac{2}{3},\tfrac{2}{3})$, $\vec{g}=(\tfrac{1}{4},\tfrac{3}{4})$, $\vec{h}=(\tfrac{3}{4},\tfrac{1}{4})$.

\item[(6-5)] One option is
\begin{center}
\begin{tabular}{ |c!{\boldv} c|c|c| c|  } 
\hline
\cplus $\oplus$ & \ccb{e}  & \ccg{f} & \ccy{g}  & \ccr{h} \\
\boldh
\ccb{e} & \ccg{f} & \ccone & \ccnot & \ccnot   \\
\hline
\ccg{f} & \ccone & \ccnot & \ccnot & \ccnot    \\
\hline
\ccy{g} & \ccnot & \ccnot   & \ccone & \ccnot  \\
\hline
\ccr{h} & \ccnot & \ccnot   & \ccnot  & \ccone \\
\hline
\end{tabular}
\end{center}
There are two self-complementary elements and it is therefore not a weak quantum effect algebra by Prop. \ref{prop:not-quantum}.
A complex model for this effect algebra is the following. 
Denote $\omega=e^{i 2 \pi /3}$.
Choose $\enul=1$, $\eone=64$, $e=4\omega$, $f=16\omega^2$, $g=8$, $h=-8$.

\item[(6-6)] One option is
\begin{center}
\begin{tabular}{|c!{\boldv} c|c|c| c|   } 
\hline
\cplus $\oplus$ & \ccb{e}  & \ccg{f} & \ccy{g}  & \ccr{h} \\
\boldh
\ccb{e} & \ccg{f} & \ccone & \ccnot & \ccnot   \\
\hline
\ccg{f} & \ccone & \ccnot & \ccnot & \ccnot    \\
\hline
\ccy{g} & \ccnot & \ccnot & \ccr{h}  & \ccone   \\
\hline
\ccr{h} & \ccnot & \ccnot  & \ccone & \ccnot  \\
\hline
\end{tabular}
\end{center}
It follows from the sum table that $e\oplus e \oplus e = \eone$ and $g \oplus g \oplus g = \eone$.
Since $e\neq g$, we conclude from Prop. \ref{prop:not-quantum} that this table cannot correspond to a weak quantum effect algebra.
A complex model for this effect algebra is the following. 
Denote $\omega=e^{i 2 \pi /3}$.
Choose $\enul=1$, $\eone=8$, $e=2\omega$, $f=4\omega^2$, $g=2\omega^2$, $h=4 \omega$.

\item[(6-7)] One option is
\begin{center}
\begin{tabular}{|c!{\boldv} c|c|c| c|   } 
\hline
\cplus $\oplus$ & \ccb{e}  & \ccg{f} & \ccy{g}  & \ccr{h} \\
\boldh
\ccb{e} & \ccnot & \ccy{g} & \ccnot & \ccone   \\
\hline
\ccg{f} & \ccy{g} & \ccr{h} & \ccone & \ccnot    \\
\hline
\ccy{g} & \ccnot & \ccone & \ccnot  & \ccnot   \\
\hline
\ccr{h} & \ccone & \ccnot  & \ccnot & \ccnot  \\
\hline
\end{tabular}
\end{center}
This is a fuzzy set effect algebra with $\vec{e}=(\tfrac{3}{4},\tfrac{1}{4})$, $\vec{f}=(\tfrac{1}{8},\tfrac{3}{8})$, $\vec{g}=(\tfrac{7}{8},\tfrac{5}{8})$, $\vec{h}=(\tfrac{1}{4},\tfrac{3}{4})$.

\item[(6-8)] One option is
\begin{center}
\begin{tabular}{|c!{\boldv} c|c|c| c|   } 
\hline
\cplus $\oplus$ & \ccb{e}  & \ccg{f} & \ccy{g}  & \ccr{h} \\
\boldh
\ccb{e} & \ccg{f} & \ccy{g} & \ccone & \ccnot   \\
\hline
\ccg{f} & \ccy{g}  & \ccone & \ccnot  & \ccnot  \\
\hline
\ccy{g} & \ccone & \ccnot  & \ccnot & \ccnot   \\
\hline
\ccr{h}  & \ccnot  & \ccnot & \ccnot & \ccone \\
\hline
\end{tabular}
\end{center}
There are two self-complementary elements and it is therefore not a quantum effect algebra.
A complex model for this effect algebra is the following. 
Choose $\enul=1$, $\eone=16$, $e=2i$, $f=-4$, $g=-8i$, $h=4$.

\item[(6-9)] One option is
\begin{center}
\begin{tabular}{|c!{\boldv} c|c|c| c|   } 
\hline
\cplus $\oplus$ & \ccb{e}  & \ccg{f} & \ccy{g}  & \ccr{h} \\
\boldh
\ccb{e} & \ccy{g} & \ccr{h} & \ccone & \ccnot   \\
\hline
\ccg{f} & \ccr{h}  & \ccy{g} & \ccnot  & \ccone  \\
\hline
\ccy{g} & \ccone & \ccnot  & \ccnot & \ccnot   \\
\hline
\ccr{h}  & \ccnot  & \ccone & \ccnot & \ccnot \\
\hline
\end{tabular}
\end{center}
We have $e\oplus e=f\oplus f$, hence by Prop. \ref{prop:not-quantum} this is not a weak quantum effect algebra.
A real model for this effect algebra is the following. 
Choose $\enul=1$, $\eone=8$, $e=2$, $f=-2$, $g=4$, $h=-4$.

\item[(6-10)] One option is
\begin{center}
\begin{tabular}{|c!{\boldv} c|c|c| c|   } 
\hline
\cplus $\oplus$ & \ccb{e}  & \ccg{f} & \ccy{g}  & \ccr{h} \\
\boldh
\ccb{e} & \ccg{f} & \ccy{g} & \ccr{h} & \ccone   \\
\hline
\ccg{f} & \ccy{g}  & \ccr{h} &  \ccone & \ccnot   \\
\hline
\ccy{g} & \ccr{h} & \ccone & \ccnot  & \ccnot    \\
\hline
\ccr{h}  & \ccone & \ccnot  & \ccnot & \ccnot \\
\hline
\end{tabular}
\end{center}
This is the scale effect algebra $\scale{6}=\{0,\tfrac{1}{5},\tfrac{2}{5},\tfrac{3}{5},\tfrac{4}{5},1 \}$. 

\end{enumerate}

\section{States}\label{sec:states}

In a physical context effects correspond to possible events of some system. 
A state is then a condition of the system, and a pair of an effect and a state determines a number that is interpreted as the probability that the effect occurs when the system is in the given state.
We recall the following mathematical definition of a state on an effect algebra.

\begin{definition}
Let $\ea$ be an effect algebra.
 A \emph{state} on $\ea$ is a function $\sigma:\ea \to [0,1]\subset \real$ such that 
 \begin{itemize}
 \item[(ST1)] $\sigma(\eone)=1$
 \item[(ST2)] $\sigma(e \oplus f) = \sigma(e) + \sigma(f)$  whenever $e\oplus f$ is defined. 
 \end{itemize}
 We denote by $\Sigma(\ea)$ the set of all states on $\ea$.
\end{definition}

It easily follows from these two conditions that a state $\sigma$ also satisfies 
\begin{itemize}
\item[(ST-a)] $\sigma(\enul)=0$
\item[(ST-b)] $\sigma(e^\bot)=1-\sigma(e)$
\end{itemize}

It is clear that the set $\Sigma(\ea)$ is convex: if $\sigma_1,\sigma_2\in\Sigma(\ea)$, then $t \sigma_1 + (1-t) \sigma_2 \in \Sigma(\ea)$ for any $t\in [0,1]$.
Therefore, the set $\Sigma(\ea)$ is either empty, a singleton set or contains infinitely many states.
The total number of states is not as important as the following two properties.

\begin{definition}
A subset $S \subseteq \Sigma(\ea)$ is
\begin{itemize}
\item \emph{separating} if $\sigma(e)=\sigma(f)$ for all $\sigma\in S$ implies $e=f$.
\item \emph{order-determining} if $\sigma(e)\leq \sigma(f)$ for all $\sigma\in S$ implies $e\leq f$.
\end{itemize}
\end{definition}

Clearly, an order-determining set of states is separating.  
However, the converse is not true; this is shown in Example \ref{ex:states5(3)} below.

\begin{example}(\emph{The unique state on $\sea{n}$}.)
Let $\sigma$ be a state on a scale effect algebra $\sea{n}$.
We have
\begin{equation*}
1=\sigma(1)=\sigma(\oplus_{n-1} \tfrac{1}{n-1} ) = (n-1) \sigma(\tfrac{1}{n-1}) \, ,
\end{equation*}
hence $\sigma(\tfrac{1}{n-1})=\tfrac{1}{n-1}$.
The other elements in $\sea{n}$ are of the form $\tfrac{k}{n-1}$ and it follows that $\sigma(\tfrac{k}{n-1})=\tfrac{k}{n-1}$.
Therefore, the unique state on $\sea{n}$ is the identity map.
The state is clearly order-determining.
\end{example}

\begin{example}(\emph{The unique state on $\dea{n}$}.)
Let $\sigma$ be a state on a sparse effect algebra $\dea{n}$.
For every $e\in\dea{n} \smallsetminus \{ \enul,\eone \}$, we have
\begin{equation*}
1=\sigma(\eone)=\sigma(e \oplus e ) = 2 \sigma(e) \, ,
\end{equation*}
hence $\sigma(e)=\tfrac{1}{2}$.
Therefore, the unique state on $\dea{n}$ is constant on all nontrivial effects.
It follows that $\Sigma(\dea{n})$ is not separating when $n \geq 4$.
\end{example}

\begin{example}(\emph{The states on $\pea{n}$}.)
Let $\sigma$ be a state on an effect algebra $\pea{n}$, $n\geq 4$.
With the notation of Sec. \ref{sec:non-inv}, we have $\sigma(e_j)=r_j$ and $\sigma(e_{j}^\bot)=1-r_j$ for every $j=1,\ldots,\ell$, where $0\leq r_j \leq 1$ and $n=2\ell +2$.
There are no other constraints, hence we conclude that $\Sigma(\pea{n})$ can be identified with $[0,1]^{\ell}$.
This set is order-determining.
\end{example}

In the next section we discuss states of fuzzy set effect algebras.
In the following we focus on effect algebras with 5 or 6 elements that are not fuzzy set effect algebras nor covered in previous examples.

\begin{example}\label{ex:states5(3)}(\emph{The states on $\ea_{5(3)}$}.)
Let $\sigma$ be a state on $\ea_{5(3)}$.
As $f\oplus f = \eone$, we have $\sigma(f)=1/2$.
Further, $e \oplus e \oplus e = \eone$ and hence $\sigma(e)=1/3$.
It follows that $\sigma(g)=2/3$.
We conclude that there is only one state, which is separating.
The unique state $\sigma$ is not order-determining as $\sigma(e) < \sigma(f)$ but $e \nleq f$.
\end{example}

\begin{example}(\emph{The states on $\ea_{6(3)}$}.)
Let $\sigma$ be a state on $\ea_{6(3)}$.
The effects $e$ and $f$ are self-complementary, hence $\sigma(e)=\sigma(f)=1/2$.
The value $\sigma(g)$ is arbitary, and then $\sigma(h)=1-\sigma(g)$.
We conclude that there are infinitely many states, but the state space is not separating.
\end{example}

\begin{example}(\emph{The states on $\ea_{6(5)}$}.)
Let $\sigma$ be a state on $\ea_{6(5)}$.
The effects $g$ and $h$ are self-complementary, hence $\sigma(g)=\sigma(h)=1/2$.
Since $e \oplus e \oplus e=\eone$, we have $\sigma(e)=1/3$.
Finally, $e \oplus f = \eone$, thus $\sigma(f)=2/3$.
We conclude that there is one state, which is not separating.
\end{example}

\begin{example}(\emph{The states on $\ea_{6(6)}$}.)
Let $\sigma$ be a state on $\ea_{6(6)}$.
We have 
\begin{equation*}
\sigma(e) + \sigma(e) = \sigma(e \oplus e) = \sigma(f)
\end{equation*}
and
\begin{equation*}
\sigma(e) + \sigma(f) = \sigma(e \oplus f) = \sigma(\eone)=1 \, .
\end{equation*}
Therefore, $\sigma(e) = 1/3$ and $\sigma(f)=2/3$.
Similarly,  $\sigma(g) = 1/3$ and $\sigma(h)=2/3$.
Hence, there is only one state and it is not separating.
\end{example}

\begin{example}(\emph{The states on $\ea_{6(8)}$}.)
Let $\sigma$ be a state on $\ea_{6(8)}$.
We have $\sigma(f)=1/2$ and $\sigma(h)=1/2$. 
It follows that $\sigma(e)=1/4$ and $\sigma(g)=3/4$.
Hence, there is only one state and it is not separating.
\end{example}

\begin{example}(\emph{The states on $\ea_{6(9)}$}.)
Let $\sigma$ be a state on $\ea_{6(9)}$.
We have 
\begin{equation*}
\sigma(e\oplus e\oplus e) = \sigma(f \oplus f \oplus f) = \sigma(\eone)=1 \, ,
\end{equation*}
hence $\sigma(e)=\sigma(f)=1/3$.
It further follows that $\sigma(g)=\sigma(h)=2/3$.
Therefore, there is only one state and it is not separating.
\end{example}

Finally, we recall that it was shown in \cite{binczak2023matrix} that all effect algebras with 2--8 elements have at least one state.
Intriguingly, a 9 element effect algebra with no states has been presented in \cite{rievcanova2001proper}.
The description of this effect algebra via effect sum table is given below.

\begin{center}
\begin{tabular}{|c!{\boldv} c|c|c|c|c|c|c|    } 
\hline
\cplus $\oplus$ & \ccv{e}  & \ccb{f}  & \ccc{g} & \ccg{h}  & \ccy{i} & \cco{j}  & \ccr{k} \\
\boldh
\ccv{e} & \ccg{h} & \cco{j} & \ccr{k} & \ccone & \ccnot & \ccnot  & \ccnot \\
\hline
\ccb{f} & \cco{j}  & \ccy{i} & \ccg{h}  & \ccnot & \ccr{k} & \ccnot & \ccone  \\
\hline
\ccc{g} & \ccr{k} & \ccg{h} & \cco{j}  & \ccnot & \ccnot  & \ccone & \ccnot \\
\hline
\ccg{h}  & \ccone  & \ccnot & \ccnot & \ccnot & \ccnot  & \ccnot & \ccnot  \\
\hline
\ccy{i}   & \ccnot  & \ccr{k} & \ccnot & \ccnot & \ccone  & \ccnot & \ccnot \\
\hline
\cco{j}   & \ccnot  & \ccnot & \ccone & \ccnot & \ccnot   & \ccnot & \ccnot \\
\hline
\ccr{k}  & \ccnot  & \ccone & \ccnot & \ccnot & \ccnot   & \ccnot & \ccnot \\
\hline
\end{tabular}
\end{center}
As $e\oplus e \oplus e = g\oplus g \oplus g$, this is not a weak quantum effect algebra by Prop. \ref{prop:not-quantum}.

\section{Finite quantum effect algebras}\label{sec:quantum}

The following results relates the state space structure to a representation of the effect algebra as a fuzzy set effect algebra.
Importantly, this leads to the conclusion that finite quantum effect algebras are the same as finite fuzzy set algebras. 

\begin{theorem}\label{thm:fse}
Let $\ea$ be an effect algebra.
The following statements are equivalent:
\begin{itemize}
\item[(i)] $\ea$ has a finite set of order-determining states.
\item[(ii)] $\ea$ is a fuzzy set effect algebra.
\end{itemize}
If $\ea$ is finite, then the following is equivalent with the previous two statements:
\begin{itemize}
\item[(iii)] $\ea$ is a quantum effect algebra.
\end{itemize}
\end{theorem}

\begin{proof}
(i)$\Rightarrow$(ii): Suppose $\ea$ is an effect algebra with a finite set of order-determining states $S$.
We define $\Phi:\ea \to [0,1]^S$ by $\Phi(e)(\sigma)=\sigma(e)$.
We have $\Phi(\eone)(\sigma)=\sigma(\eone)=1$, hence $\Phi(\eone)=1$.
Further, we have
\begin{equation*}
\Phi(e^\bot)(\sigma)=\sigma(e^\bot)=1-\sigma(e) = 1 - \Phi(e)(\sigma) = \Phi(e)^\bot(\sigma) \, ,
\end{equation*}
hence $\Phi(e^\bot)=\Phi(e)^\bot$.
If $e,f\in \ea$ with $e\oplus f$ defined, then
\begin{align*}
\Phi(e \oplus f)(\sigma) & =\sigma(e \oplus f) = \sigma(e) + \sigma(f)= \Phi(e)(\sigma) + \Phi(f)(\sigma) \\
& = [\Phi(e) + \Phi(f)](\sigma) \, .
\end{align*}
Hence, $\Phi(e \oplus f) = \Phi(e) + \Phi(f)$, so $\Phi(e) \oplus \Phi(f)$ is defined in $[0,1]^S$ and $\Phi(e) \oplus \Phi(f)=\Phi(e) + \Phi(f)$.
We conclude that $\Phi$ is a morphism.
To show that $\Phi$ is a monomorphism, suppose that $\Phi(e) \oplus \Phi(f)$ is defined.
Then for all $\sigma \in S$, we have
\begin{equation*}
\sigma(e) + \sigma(f) = \Phi(e)(\sigma) + \Phi(f)(\sigma) = [\Phi(e) + \Phi(f)](\sigma)= [\Phi(e) \oplus \Phi(f)](\sigma) \leq 1 \, ,
\end{equation*}
which means that $\sigma(e) \leq 1- \sigma(f) = \sigma(f^\bot)$.
This is valid for all $\sigma \in S$ and since $S$ is order-determining, $e\leq f^\bot$ and hence $e \oplus f$ is defined.
 and $\Phi(e \oplus f)=\Phi(e) \oplus \Phi(f)$.
This means that $\Phi$ is a monomorphism and thus $\Phi:\ea \to \Phi(\ea)$ is an isomorphism.
Since $\Phi(\ea)$ is an effect subalgebra of $[0,1]^S$, we conclude that $\ea$ is a fuzzy set effect algebra.

(ii)$\Rightarrow$(i): Suppose $\ea$ is a fuzzy set effect algebra.
We can assume $\ea \subset [0,1]^n$.
For each $k=1,\ldots,n$, we define $\sigma_k:[0,1]^n\to[0,1]$ as $\sigma_k(\vx)=x_k$.
One can confirm that $\sigma_k$ is a state and we denote by $S$ the set of all these states.
Assume $\vx,\vy\in\ea$ with $\sigma(\vx)\leq \sigma(\vy)$ for all $\sigma\in S$.
Then $x_k \leq y_k$ for all $k=1,\ldots,n$, so that $x_k + (1-y_k) \leq 1$. 
Hence, $\vx \oplus \vy^\bot$ is defined, which means that also $(\vx \oplus \vy^\bot)^\bot$ is defined.
Thus, by (EA-e) we have
\begin{equation*}
\vy \ominus \vx = (\vx \oplus \vy^\bot)^\bot \in \ea \, .
\end{equation*}
Therefore, $\vx \oplus (\vy \ominus \vx) = \vy$, so that $\vx \leq \vy$.
We conclude that $S$ is order-determining.

To prove the final claim, let $\ea$ be a finite effect algebra. To show that (iii) is equivalent to (i) and (ii), it is enough to show that (iii)$\Rightarrow$(i) as (ii)$\Rightarrow$(iii) by Prop. \ref{prop:anyis}.
A finite quantum effect algebra $\ea=\{0,\id,E_1,\ldots,E_n \}$ is a subset of $\eh$ and $E_i\oplus E_j$ is defined if and only if $E_i + E_j \leq \id$, in which case  $E_i\oplus E_j=E_i + E_j$.
If $E_i,E_j\in \ea$ are such that $E_i \nleq E_j$ in the ordering of $\ea$, then by Prop. \ref{prop:order}b $E_i \nleq E_j$ in the usual operator ordering. 
Hence, there exists a unit vector $\psi_{ij}\in\hi$ such that $\ip{\psi_{ij}}{E_i\psi_{ij}} > \ip{\psi_{ij}}{ E_j \psi_{ij}}$.
We define a state $\sigma_{ij}$ as $\sigma_{ij} = \ip{\psi_{ij}}{E\psi_{ij}}$. 
The set of states
$$
S = \{ \sigma_{ij} : E_i \nleq E_j \}
$$
is finite and order-determining.
\end{proof}

A finite quantum effect algebra of dimension $d$ may not be a fuzzy set effect algebra of the same dimension $d$.
For instance, let $\ea_8$ be the effect algebra corresponding to the following sum table.

\begin{center}
\begin{tabular}{|c!{\boldv} c|c|c|c|c|c|    } 
\hline
\cplus $\oplus$ & \ccv{e}  & \ccb{f} & \cg  & \ch & \ci & \cj \\
\boldh
\ce & \ccnot & \ch & \ci & \ccnot & \ccnot & \ccone  \\
\hline
\ccb{f} & \ch  & \ccnot & \cj  & \ccnot & \ccone & \ccnot  \\
\hline
\cg & \ci & \cj & \ccnot  & \ccone & \ccnot  & \ccnot  \\
\hline
\ch  & \ccnot  & \ccnot & \ccone & \ccnot & \ccnot  & \ccnot  \\
\hline
\ci  & \ccnot  & \ccone & \ccnot & \ccnot & \ccnot  & \ccnot \\
\hline
\cj  & \ccone  & \ccnot & \ccnot & \ccnot & \ccnot   & \ccnot \\
\hline
\end{tabular}
\end{center}
This effect algebra $\ea_8$ is generated by the subset $\{e,f,g\}$ and the generators satisfy $e \oplus f \oplus g = \eone$.
We make the following observations.
\begin{itemize}
\item $\ea_8$ is a fuzzy set effect algebra of dimension 3 with $\vec{e}=(1,0,0)$, $\vec{f}=(0,1,0)$, $\vec{g}=(0,0,1)$.
\item $\ea_8$ is a not a fuzzy set effect algebra of dimension 2.
To see this, assume on the contrary that $\vec{e}=(e_1,e_2)$, $\vec{f}=(f_1,f_2)$, $\vec{g}=(g_1,g_2)$.
From  $e \oplus f \oplus g = \eone$ follows that
\begin{align*}
e_1+f_1+g_1 = 1 \\
e_2+f_2+g_2 = 1
\end{align*}
This means that at least two numbers among $e_1,f_1,g_1$ are at most $\half$, and similarly at least two numbers among $e_2,f_2,g_2$ are at most $\half$.
It follows that at least one of the vectors $\vec{e},\vec{f},\vec{g}$ has both components at most $\half$; suppose it is $\vec{e}$.
Then $\vec{e}+ \vec{e} \leq (1,1)$. But $\vec{e} \oplus \vec{e}$ is not defined, hence a contradiction.
\item $\ea_8$ is a quantum effect algebra of dimension 2. 
One possible choice is to take
\begin{equation*}
E = \frac{1}{3} \left(\begin{array}{cc}2 & 0 \\0 & 0\end{array}\right) \, , \  F = \frac{1}{6} \left(\begin{array}{cc}1 & \sqrt{3} \\ \sqrt{3} &  3\end{array}\right) \, , \ G = \frac{1}{6} \left(\begin{array}{cc}1 & -\sqrt{3} \\ -\sqrt{3} & 3\end{array}\right) \, .
\end{equation*}
The eigenvalues of each of these matrices are $0$ and $\tfrac{2}{3}$.
Hence, they are quantum effects.
It is a straightforward calculation to check that they satisfy the sum table of $\ea_8$.
\end{itemize}

We conclude that for finite effect algebras it is the same to be a fuzzy set effect algebra than being a quantum effect algebra.
However, the key difference is manifested in the dimensions of the representations, which can vary.

\section{Composition}\label{sec:composition}

Let $\ea_1$ and $\ea_2$ be two effect algebras.
Their Cartesian product $\ea_1 \times \ea_2$ becomes an effect algebra when the effect sum is defined componentwise:
for $(e_1,e_2),(f_1,f_2)$ we define 
$$
(e_1,e_2)\oplus (f_1,f_2) = (e_1\oplus f_1, e_2 \oplus f_2)
$$
 if both $e_1\oplus f_1$ and $e_2 \oplus f_2$ are defined; otherwise $(e_1,e_2)\diamond (f_1,f_2)$. 
One can then verify that the conditions (EA1) -- (EA4) are satisfied with $\enul=(\enul_1,\enul_2)$ and $\eone=(\eone_1,\eone_2)$.
We call $\ea_1 \times \ea_2$ the \emph{composite effect algebra} for $\ea_1$ and $\ea_2$.
We also say that $\ea$ is a composite effect algebra if $\ea=\ea_1 \times \ea_2$ for some $\ea_1$ and $\ea_2$.

Given a composite effect algebra $\ea_1 \times \ea_2$, we see that $\ea_1$ is isomorphic to the effect algebra
\begin{equation*}
\hat{\ea}_1 =  \{ (e,\enul_2) : e \in \ea_1\} \, ,
\end{equation*}
and we call it the \emph{first component} of $\ea_1 \times \ea_2$.
The identity element of $\hat{\ea}_1$ is $(\eone_1, \enul_2)$, hence it is not an effect subalgebra of $\ea_1 \times \ea_2$.
In a similar way, the effect algebra
\begin{equation*}
\hat{\ea}_2 =  \{ (\enul_1,e) : e \in \ea_2\}
\end{equation*}
is  the \emph{second component} of $\ea_1 \times \ea_2$ and it is isomorphic to $\ea_2$.

\begin{proposition}
The composite effect algebras $\ea_1 \times \ea_2$ and $\ea_2 \times \ea_1$ are isomorphic.
\end{proposition}

\begin{proof}
We define a map $\Phi:\ea_1 \times \ea_2 \to \ea_2 \times \ea_1$ as $\Phi[(e_1,e_2)] = (e_2,e_1)$.
Then $\Phi(\eone_1,\eone_2)]=(\eone_2,\eone_1)$ and 
\begin{align*}
\Phi[(e_1,e_2)\oplus (f_1,f_2)] & = \Phi[(e_1\oplus f_1,e_2\oplus f_2)] = (e_2\oplus f_2,e_1\oplus f_1) \\
& = (e_2,e_1) \oplus (f_2,f_1) = \Phi[(e_1,e_2)] \oplus \Phi[(f_1,f_2)] \, .
\end{align*}
If $\Phi[(e_1,e_2)] \oplus \Phi[(f_1,f_2)]$ is defined, then $(e_2,e_1) \oplus (f_2,f_1)$ is defined so $e_1 \oplus f_1$ and $e_2 \oplus f_2$ are defined.
Hence, $(e_1,e_2)\oplus (f_1,f_2)$ is defined and we conlude that
\begin{align*}
\Phi[(e_1,e_2)\oplus (f_1,f_2)] = \Phi[(e_1,e_2)] \oplus \Phi[(f_1,f_2)] \, .
\end{align*}
Thus, $\Phi$ is a monomorphism. Since $\Phi$ is surjective, it is an isomorphism.
\end{proof}

The simplest composite effect algebras are compositions of small scale effect algebras $\scale{2}$, $\scale{3}$ and $\scale{4}$. 
Before we take a look on some examples, we make the following simple observation.  

\begin{proposition}
A scale effect algebra $\scale{n}$ is not a composite effect algebra.
\end{proposition}

\begin{proof}
Let $\ea_1,\ea_2$ be two effect algebras.
The elements $(\eone_1, \enul_2)$ and $(\enul_1, \eone_2)$ of $\ea_1 \times \ea_2$ are not comparable.
By Theorem \ref{thm:scale}, $\scale{n}$ is totally ordered, hence, $\scale{n} \neq \ea_1 \times \ea_2$.
\end{proof}

In particular, it follows that the composition of two scale effect algebras is not a scale effect algebra.
In the following we identify the smallest composite effect algebras.

\begin{example}\label{ex:not-scale}
The simplest composite effect algebra is $\scale{2}\times\scale{2}$.
By calculating the sum table one confirms that this is the effect algebra $\pea{4}$.
As we have seen earlier, $\pea{4}$ is a fuzzy set effect algebra but not a scale effect algebra.
We hence conclude that a composition of two scale effect algebras need not be a scale effect algebra.
Since there are no other effect algebras with two elements, we conclude that among effect algebras of order 4, the only composite effect algebra is $\pea{4}$.
\end{example}

\begin{example}
The next simplest composite effect algebra is $\scale{2}\times\scale{3}$.
By calculating the sum table one confirms that this is the effect algebra $\ea_{6(7)}$, which is a 2-dimensional fuzzy set effect algebra.
Since there are no other effect algebras with two and three elements, we conclude that among effect algebras of order 6, the only composite effect algebra is $\ea_{6(7)}$.
\end{example}

\begin{example}
The composite effect algebra $\scale{2}\times\scale{4}$ has 8 elements and the sum table is the following.
\begin{center}
\begin{tabular}{|c!{\boldv} c|c|c|c|c|c|    } 
\hline
\cplus $\oplus$ & \ccv{e}  & \ccb{f} & \ccg{g}  & \ccy{h} & \cco{i} & \ccr{j} \\
\boldh
\ccv{e} & \ccb{f} & \ccg{g} & \ccnot & \cco{i} & \ccr{j} & \ccone  \\
\hline
\ccb{f} & \ccg{g}  & \ccnot & \ccnot  & \ccr{j} & \ccone & \ccnot  \\
\hline
\ccg{g} & \ccnot & \ccnot & \ccnot  & \ccone & \ccnot  & \ccnot  \\
\hline
\ccy{h}  & \cco{i}  & \ccr{j} & \ccone & \ccnot & \ccnot  & \ccnot  \\
\hline
\cco{i}  & \ccr{j}  & \ccone & \ccnot & \ccnot & \ccnot  & \ccnot \\
\hline
\ccr{j}  & \ccone  & \ccnot & \ccnot & \ccnot & \ccnot   & \ccnot \\
\hline
\end{tabular}
\end{center}
This is not isomorphic with $\ea_8$ presented in Sec. \ref{sec:quantum} as they have different number of undefined pairs.
\end{example}

Let us then consider states on composite effect algebras.
Let $\sigma_1 \in \Sigma(\ea_1)$, $\sigma_2 \in \Sigma(\ea_2)$ and $t \in [0,1]$.
We define a map $t \sigma_1 \otimes (1-t)\sigma_2$ on $\ea_1 \times \ea_2$ by 
\begin{equation}
\left( t \sigma_1 \otimes (1-t)\sigma_2 \right) [(e_1,e_2)] =  t \sigma_1(e_1) + (1-t)\sigma_2(e_2) \, . 
\end{equation}

\begin{proposition}\label{prop:tstates}
The map $t \sigma_1 \otimes (1-t)\sigma_2$ is a state on $\ea_1 \times \ea_2$. The states of this form are called product states.
\end{proposition}

\begin{proof}
Denote $\sigma=t \sigma_1 \otimes (1-t)\sigma_2$.
We have that
\begin{equation*}
\sigma [(\eone,\eone)] = t \sigma_1(\eone) + (1-t)\sigma_2(\eone)=1 \, .
\end{equation*}
If $(e_1,e_2),(f_1,f_2)\in \ea_1 \times \ea_2$ with $(e_1,e_2)\oplus(f_1,f_2)$ defined, we obtain
\begin{align*}
& \sigma [(e_1,e_2)\oplus (f_1,f_2)]  = \sigma [(e_1\oplus f_1,e_2 \oplus f_2)] \\
&=  t \sigma_1 (e_1 \oplus f_1) + (1-t) \sigma_2 (e_2 \oplus f_2) \\
& =  t \sigma_1 (e_1) + t \sigma_1(f_1) + (1-t) \sigma_2 (e_2)  + (1-t) \sigma_2(f_2) \\ 
&= \sigma [(e_1,e_2)] + \sigma[ (f_1,f_2)] 
\end{align*}
Hence, $t \sigma_1 \otimes (1-t)\sigma_2 \in \Sigma(\ea_1 \times \ea_2)$.
\end{proof}

Let $\sigma$ be a state on $\ea_1 \times \ea_2$. 
We define a map $\sigma_{[1]}$ on $\ea_1$ by
\begin{equation}
\sigma_{[1]}(e)=\sigma[(e,\enul_2)] / \sigma[(\eone_1,\enul_2)]\, ,
\end{equation}
and in a similar way we define $\sigma_{[2]}$ on $\ea_2$,
\begin{equation}
\sigma_{[2]}(e)=\sigma[(\enul_1,e)] / \sigma[(\enul_1,\eone_2)]\, .
\end{equation}
In these definitions we have to assume that $\sigma[(\eone_1,\enul_2)] \neq 0$ and $ \sigma[(\enul_1,\eone_2)] \neq 0$, respectively.

\begin{proposition}
The map $\sigma_{[i]}$ is a state on $\ea_i$ for $i=1,2$.
\end{proposition}

\begin{proof}
We have
\begin{equation}
\sigma_{[1]}(\eone_1)=\sigma[(\eone_1,\enul_2)] / \sigma[(\eone_1,\enul_2)] = 1 \, ,
\end{equation}
hence (ST1) holds. 
Let $e,f \in \ea_1$ such that $e \oplus f$ is defined.
Then
\begin{equation}
\sigma_{[1]}(e \oplus f)=\frac{\sigma[(e \oplus f,\enul_2)]}{ \sigma[(\eone_1,\enul_2)]} = \frac{\sigma[(e ,\enul_2)] + \sigma[(f,\enul_2)]}{ \sigma[(\eone_1,\enul_2)]} =  \sigma_{[1]}(e) + \sigma_{[1]}( f)\, ,
\end{equation}
so that (ST2) holds.
\end{proof}

The previous two constructions are linked in the following way.

\begin{proposition}
Let  $\sigma=t \sigma_1 \otimes (1-t)\sigma_2$ for some $\sigma_1 \in \Sigma(\ea_1)$, $\sigma_2 \in \Sigma(\ea_2)$ and $0<t <1$.
Then $\sigma_{[1]}=\sigma_1$ and $\sigma_{[2]}=\sigma_2$.
\end{proposition}

\begin{proof}
First, we need to check that $\sigma[(\eone_1,\enul_2)] \neq 0$ and  $ \sigma[(\enul_1,\eone_2)] \neq 0$.
These inequalities hold as
\begin{equation*}
\sigma[(\eone_1,\enul_2)] = \left( t \sigma_1 \otimes (1-t)\sigma_2 \right) [(\eone_1,\enul_2)] =  t \sigma_1(\eone_1) + (1-t)\sigma_2(\enul_2) =t
\end{equation*}
and
\begin{equation*}
\sigma[(\enul_1,\eone_2)] = \left( t \sigma_1 \otimes (1-t)\sigma_2 \right) [(\enul_1,\eone_2)] =  t \sigma_1(\enul_1) + (1-t)\sigma_2(\eone_2) =1-t \,.
\end{equation*}
We then have
\begin{equation*}
\sigma_{[1]}(e)=\frac{\sigma[(e,\enul_2)]}{\sigma[(\eone_1,\enul_2)]}=\frac{t \sigma_1(e) + (1-t)\sigma_2(\enul_2)}{t} = \sigma_1(e)
\end{equation*}
and
\begin{equation*}
\sigma_{[2]}(e)=\frac{\sigma[(\enul_1,e)]}{\sigma[(\enul_1,\eone_2)]}=\frac{t \sigma_1(\enul_1) + (1-t)\sigma_2(e)}{1-t} = \sigma_2(e) \, .
\end{equation*}
\end{proof}

We have seen in Example \ref{ex:not-scale} that the composition of two scale effect algebras need not be a scale effect algebra.
In contrast, the composition of two fuzzy set effect algebras is a fuzzy set effect algebra.
In fact, the class of fuzzy set effect algebras is closed under composition in the following strong sense.

\begin{proposition}
Let $\ea_1$ and $\ea_2$ be effect algebras.
The composite effect algebra $\ea_1 \times \ea_2$ is a fuzzy set effect algebra if and only if both $\ea_1$ and $\ea_2$ are fuzzy set effect algebras.
\end{proposition}

\begin{proof}
Suppose that $\ea_1$ and $\ea_2$ are fuzzy set effect algebras.
By Theorem \ref{thm:fse}, $\ea_1$ and $\ea_2$ have finite sets of order-determining states, $S_1$ and $S_2$, respectively.
Let
\begin{equation*}
S= \{ \sigma'_1, \sigma'_2 : \sigma_1 \in S_1, \sigma_2 \in S_2 \} \, , 
\end{equation*}
where $\sigma'_i(e_1,e_2)=\sigma_i(e_i)$ for $i=1,2$.
By Prop. \ref{prop:tstates}, the elements of $S$ are states on $\ea_1 \times \ea_2$.
To show that $S$ is order-determining, suppose $\sigma[(e_1,e_2)] \leq \sigma[(f_1,f_2)]$ for all $\sigma \in S$.
It follows that $\sigma_i(e_i)\leq \sigma_i(f_i)$ for all $\sigma_i \in S_i$ with $i=1,2$.
Since $S_1$ and $S_2$ are order-determining, we have $e_1 \leq f_1$ and $e_2 \leq f_2$.
Therefore, $(e_1,e_2) \leq (f_1,f_2)$, and we conclude that $S$ is order-determining.
By Theorem \ref{thm:fse}, $\ea_1 \times \ea_2$ is a fuzzy set effect algebra.

Suppose then that $\ea_1 \times \ea_2$ is a fuzzy set effect algebra.
Let $S$ be a finite order-determining set of states on $\ea_1 \times \ea_2$ and $S_1$ the subset of $S$ that satisfy  $\sigma[(\eone_1,\enul_2)] \neq 0$.
This subset is order determining on the subsets of effects of the form $(e,\enul_2)$, as the other states are identically zero on this subset.
Let $e,f\in\ea_1$.
Every $\sigma \in S_1$ defines a state $\sigma_{[1]}$ on $\ea_1$.
Assume that $\sigma_{[1]}(e) \leq \sigma_{[1]}(f)$ for all $\sigma \in S_1$.
It follows that $\sigma[(e,\enul_2)] \leq \sigma[(f,\enul_2)]$ for all $\sigma \in S_1$, and therefore $e \leq f$.
This shows that $\{ \sigma_{[1]} : \sigma \in S_1 \}$ is a finite order determining set of states on $\ea_1$. Hence, by Theorem \ref{thm:fse},  $\ea_1$ is a fuzzy set effect algebra. In a similar way, $\ea_2$ is a fuzzy set effect algebra.
\end{proof}

\section{Conclusions}

We have introduced sum tables as an illustrative tool to describe finite effect algebras. Determining whether a table defines a valid effect algebra is not immediately obvious; however, we have provided a set of criteria that all sum tables must satisfy.
As a demonstration of the usefulness of sum tables, we showed that all effect algebras with 2–6 elements can be systematically listed by specifying their sum tables (see Appendix B).

Focusing on finite effect algebras brings forth natural questions that do not typically arise in the study of general effect algebras. In particular, one might ask about the maximal and minimal numbers of undefined effect pairs. We have characterized both of these extreme cases and uncovered the structural implications they entail.

We have shown that every finite quantum effect algebra is also a fuzzy set efffect algerba. 
Hence, in that sense `quantumness' is not immediately apparent in the effect algebra structure. 
However, as we demonstrated with an example, a finite quantum effect algebra of dimension $d$ may not be a fuzzy set effect algebra of the same dimension $d$.
This observation warrants further investigation and will be addressed in future work.

\newpage

\section*{Appendix A: Some basic properties of effect algebras}

\begin{proposition}
Let $e,f,g \in \ea$.
\begin{itemize}
\item[(a)] $e \oplus \enul = e$.
\item[(b)] If $e \oplus f = \enul$, then $e=f=\enul$.
\item[(c)] $e \ominus f = (e^\bot \oplus f)^\bot$.
\item[(d)] $e \oplus f$ is defined if and only if $e\leq f^\bot$. 
\item[(e)] $(e \oplus f) \ominus e = f$.
\item[(f)] If $e \oplus f = e \oplus g$, then $f=g$.
\item[(g)] If $e \oplus f = e$, then $f=\enul$.

\end{itemize}
\end{proposition}


\begin{proof}
\begin{itemize}
\item[(a)] Since $\enul \oplus (e \oplus e^\bot)$ is defined, also $(\enul \oplus e) \oplus e^\bot$ is defined and hence
\begin{equation*}
(\enul \oplus e) \oplus e^\bot = \enul \oplus (e \oplus e^\bot) = \enul \oplus \eone = \eone \, .
\end{equation*}
It follows from (EA3) that $\enul \oplus e = e$.
\item[(b)] Suppose $e \oplus f = \enul$.  Then $e^\bot \oplus (e \oplus f)$ is defined and therefore
\begin{equation*}
e^\bot \oplus (e \oplus f) = (e^\bot \oplus e) \oplus f = \eone \oplus f \, .
\end{equation*}
It follows from (EA4) that $f=\enul$. In a similar way we can deduce that $e=\enul$.
\item[(c)] Let $e=f \oplus g$. Then
\begin{equation*}
\eone = e^\bot \oplus e = e^\bot \oplus (f \oplus g) = (e^\bot \oplus f) \oplus g \, ,
\end{equation*}
hence $g^\bot = e^\bot \oplus f$. Taking the complement on both sides gives $g = (e^\bot \oplus f)^\bot$.
\item[(d)] Suppose $e \oplus f = e$. Then $e^\bot \oplus (e \oplus f)$ is defined and we get
\begin{equation*}
e^\bot \oplus (e \oplus f) =( e^\bot \oplus e) \oplus f = \eone \oplus f \, .
\end{equation*}
It follows from (EA4) that $f=\enul$.
\item[(e)] Suppose $e \oplus f$ is defined. Since $(e \oplus f)^\bot \oplus (e \oplus f)=\eone$, we have $(e \oplus f)^\bot \oplus e = f^\bot$. Applying (c) gives 
\begin{equation*}
(e \oplus f) \ominus e = ((e \oplus f)^\bot \oplus e)^\bot = f \, .
\end{equation*}
\item[(f)] Suppose $e \oplus f = e \oplus g$. Applying (e) we get
\begin{equation*}
f =  (e \oplus f) \ominus e = (e \oplus g) \ominus e = g \, .
\end{equation*}
\item[(g)] This follows from (a) and (f).
\end{itemize}
\end{proof}

\begin{landscape}

\section*{Appendix B: The sum tables of effect algebras with 3 -- 6 elements.}

\

\begin{center}
\begin{tabular}{|c!{\boldv}c| } 
\hline
\cplus $\oplus$ & \ccb{e}  \\
\boldh
\ccb{e} & \ccone  \\
\hline
\end{tabular}
\end{center}

\

\begin{center}
\begin{tabular}{|c!{\boldv}c|c| } 
\hline
\cplus $\oplus$ & \ccb{e}  & \ccr{f} \\
\boldh
\ccb{e} & \ccone & \ccnot   \\
\hline
\ccr{f} & \ccnot & \ccone    \\
\hline
\end{tabular} \begin{tabular}{|c!{\boldv}c|c| } 
\hline
\cplus $\oplus$ & \ccb{e}  & \ccr{f} \\
\boldh
\ccb{e} & \ccnot & \ccone   \\
\hline
\ccr{f} & \ccone & \ccnot    \\
\hline
\end{tabular} \begin{tabular}{|c!{\boldv}c|c| } 
\hline
\cplus $\oplus$ & \ccb{e}  & \ccr{f} \\
\boldh
\ccb{e} & \ccr{f} & \ccone   \\
\hline
\ccr{f} & \ccone & \ccnot    \\
\hline
\end{tabular}
\end{center}

\

\begin{center}
\begin{tabular}{|c!{\boldv}c|c|c|  } 
\hline
\cplus $\oplus$ & \ccb{e}  & \ccy{f} & \ccr{g} \\
\boldh
\ccb{e} & \ccone & \ccnot & \ccnot   \\
\hline
\ccy{f} & \ccnot & \ccone  & \ccnot   \\
\hline
\ccr{g} & \ccnot & \ccnot  & \ccone   \\
\hline
\end{tabular} \begin{tabular}{|c!{\boldv}c|c|c|  } 
\hline
\cplus $\oplus$ & \ccb{e}  & \ccy{f} & \ccr{g} \\
\boldh
\ccb{e} & \ccone & \ccnot & \ccnot   \\
\hline
\ccy{f} & \ccnot & \ccnot & \ccone    \\
\hline
\ccr{g} & \ccnot & \ccone & \ccnot     \\
\hline
\end{tabular} \begin{tabular}{|c!{\boldv}c|c|c|  } 
\hline
\cplus $\oplus$ & \ccb{e}  & \ccy{f} & \ccr{g} \\
\boldh
\ccb{e} & \ccr{g} & \ccnot & \ccone   \\
\hline
\ccy{f} & \ccnot & \ccone  & \ccnot   \\
\hline
\ccr{g} & \ccone & \ccnot  & \ccnot   \\
\hline
\end{tabular} \begin{tabular}{|c!{\boldv}c|c|c|  } 
\hline
\cplus $\oplus$ & \ccb{e}  & \ccy{f} & \ccr{g} \\
\boldh
\ccb{e} & \ccy{f} & \ccr{g} & \ccone   \\
\hline
\ccy{f} & \ccr{g} & \ccone  & \ccnot   \\
\hline
\ccr{g} & \ccone & \ccnot  & \ccnot   \\
\hline
\end{tabular}
\end{center}

\

\begin{center}
\begin{tabular}{ |c!{\boldv} c|c|c| c|   } 
\hline
\cplus $\oplus$ & \ccb{e}  & \ccg{f} & \ccy{g}  & \ccr{h} \\
\boldh
\ccb{e} & \ccone & \ccnot & \ccnot  & \ccnot \\
\hline
\ccg{f} & \ccnot & \ccone  & \ccnot & \ccnot   \\
\hline
\ccy{g} & \ccnot & \ccnot  & \ccone & \ccnot  \\
\hline
\ccr{h} & \ccnot & \ccnot  &  \ccnot & \ccone   \\
\hline
\end{tabular} \begin{tabular}{ |c!{\boldv} c|c|c| c|   } 
\hline
\cplus $\oplus$ & \ccb{e}  & \ccg{f} & \ccy{g}  & \ccr{h} \\
\boldh
\ccb{e} & \ccnot & \ccone & \ccnot & \ccnot   \\
\hline
\ccg{f} & \ccone & \ccnot & \ccnot & \ccnot    \\
\hline
\ccy{g} & \ccnot & \ccnot & \ccnot  & \ccone   \\
\hline
\ccr{h} & \ccnot & \ccnot   & \ccone & \ccnot  \\
\hline
\end{tabular} \begin{tabular}{ |c!{\boldv} c|c|c| c|   } 
\hline
\cplus $\oplus$ & \ccb{e}  & \ccg{f} & \ccy{g}  & \ccr{h} \\
\boldh
\ccb{e} & \ccone & \ccnot & \ccnot & \ccnot   \\
\hline
\ccg{f} & \ccnot & \ccone & \ccnot  & \ccnot  \\
\hline
\ccy{g} & \ccnot & \ccnot & \ccnot & \ccone   \\
\hline
\ccr{h} & \ccnot & \ccnot  & \ccone & \ccnot  \\
\hline
\end{tabular} \begin{tabular}{|c!{\boldv} c|c|c| c|   } 
\hline
\cplus $\oplus$ & \ccb{e}  & \ccg{f} & \ccy{g}  & \ccr{h} \\
\boldh
\ccb{e} & \ccg{f} & \ccone & \ccnot & \ccnot   \\
\hline
 \ccg{f} & \ccone & \ccnot & \ccnot & \ccnot    \\
\hline
 \ccy{g} & \ccnot & \ccnot & \ccnot  & \ccone   \\
\hline
 \ccr{h} & \ccnot & \ccnot   & \ccone & \ccnot  \\
\hline
\end{tabular} \begin{tabular}{ |c!{\boldv} c|c|c| c|  } 
\hline
\cplus $\oplus$ & \ccb{e}  & \ccg{f} & \ccy{g}  & \ccr{h} \\
\boldh
\ccb{e} & \ccg{f} & \ccone & \ccnot & \ccnot   \\
\hline
\ccg{f} & \ccone & \ccnot & \ccnot & \ccnot    \\
\hline
\ccy{g} & \ccnot & \ccnot   & \ccone & \ccnot  \\
\hline
\ccr{h} & \ccnot & \ccnot   & \ccnot  & \ccone \\
\hline
\end{tabular} 
\end{center}

\

\begin{center}
\begin{tabular}{|c!{\boldv} c|c|c| c|   } 
\hline
\cplus $\oplus$ & \ccb{e}  & \ccg{f} & \ccy{g}  & \ccr{h} \\
\boldh
\ccb{e} & \ccg{f} & \ccone & \ccnot & \ccnot   \\
\hline
\ccg{f} & \ccone & \ccnot & \ccnot & \ccnot    \\
\hline
\ccy{g} & \ccnot & \ccnot & \ccr{h}  & \ccone   \\
\hline
\ccr{h} & \ccnot & \ccnot  & \ccone & \ccnot  \\
\hline
\end{tabular} \begin{tabular}{|c!{\boldv} c|c|c| c|   } 
\hline
\cplus $\oplus$ & \ccb{e}  & \ccg{f} & \ccy{g}  & \ccr{h} \\
\boldh
\ccb{e} & \ccnot & \ccy{g} & \ccnot & \ccone   \\
\hline
\ccg{f} & \ccy{g} & \ccr{h} & \ccone & \ccnot    \\
\hline
\ccy{g} & \ccnot & \ccone & \ccnot  & \ccnot   \\
\hline
\ccr{h} & \ccone & \ccnot  & \ccnot & \ccnot  \\
\hline
\end{tabular} \begin{tabular}{|c!{\boldv} c|c|c| c|   } 
\hline
\cplus $\oplus$ & \ccb{e}  & \ccg{f} & \ccy{g}  & \ccr{h} \\
\boldh
\ccb{e} & \ccg{f} & \ccy{g} & \ccone & \ccnot   \\
\hline
\ccg{f} & \ccy{g}  & \ccone & \ccnot  & \ccnot  \\
\hline
\ccy{g} & \ccone & \ccnot  & \ccnot & \ccnot   \\
\hline
\ccr{h}  & \ccnot  & \ccnot & \ccnot & \ccone \\
\hline
\end{tabular} \begin{tabular}{|c!{\boldv} c|c|c| c|   } 
\hline
\cplus $\oplus$ & \ccb{e}  & \ccg{f} & \ccy{g}  & \ccr{h} \\
\boldh
\ccb{e} & \ccy{g} & \ccr{h} & \ccone & \ccnot   \\
\hline
\ccg{f} & \ccr{h}  & \ccy{g} & \ccnot  & \ccone  \\
\hline
\ccy{g} & \ccone & \ccnot  & \ccnot & \ccnot   \\
\hline
\ccr{h}  & \ccnot  & \ccone & \ccnot & \ccnot \\
\hline
\end{tabular} \begin{tabular}{|c!{\boldv} c|c|c| c|   } 
\hline
\cplus $\oplus$ & \ccb{e}  & \ccg{f} & \ccy{g}  & \ccr{h} \\
\boldh
\ccb{e} & \ccg{f} & \ccy{g} & \ccr{h} & \ccone   \\
\hline
\ccg{f} & \ccy{g}  & \ccr{h} &  \ccone & \ccnot   \\
\hline
\ccy{g} & \ccr{h} & \ccone & \ccnot  & \ccnot    \\
\hline
\ccr{h}  & \ccone & \ccnot  & \ccnot & \ccnot \\
\hline
\end{tabular}
\end{center}

\end{landscape}

\end{document}